\documentclass{article}

\usepackage{geometry}
\newgeometry{vmargin={20mm}, hmargin={17mm,17mm}}   

\usepackage{framed,multirow}
\usepackage{subfigure}
\usepackage[compat=1.0.0]{tikz-feynman}

\usepackage{amssymb}
\usepackage{latexsym}
\usepackage{authblk}

\usepackage{url}
\usepackage{xcolor}

\definecolor{newcolor}{rgb}{.8,.349,.1}

\newcommand{\meg}{\ensuremath{\mu^+ \to e^+ \gamma}}
\newcommand{\meee}{\ensuremath{\mu^+ \to e^+ e^+ e^-}}

\title{Experimental searches for muon decays beyond the Standard Model}%

\author[1]{Francesco Renga\footnote{Corresponding author: Email: francesco.renga@roma1.infn.it; Tel.: +39-06-4991-4266.}}

\affil[1]{Istituto Nazionale di Fisica Nucleare Ð Sez. di Roma, P.le A. Moro 2, 00185 Roma}

\begin{document}
\maketitle

\begin{abstract}
The study of muon properties and decays played a crucial role in the early years of particle physics 
and contributed over decades to build and consolidate the Standard Model. At present, searches
for muon decays beyond the Standard Model are performed by exploiting intense beams of muons,
and plans exist to upgrade the present facilities or build new ones, which would open new prospects
for the quest of new physics in this sector. In this paper I review the present status of the search
for muon decays beyond the Standard Model, with a special attention to the most conventional 
muon lepton flavor violation experiments, but also considering more exotic scenarios 
and future outlooks.
\end{abstract}





\section{Introduction}

Since the discovery of the muon in the cosmic ray radiation~\cite{muon-discovery}, the study of the properties and decays of this
particle contributed to build and test the Standard Model (SM) of particle physics. Although this particle was initially identified as the
mesotron proposed by Yukawa~\cite{mesotron} in its theory of strong interactions, this interpretation was discarded
by the experiments conducted by Conversi, Pancini and Piccioni in the years 1940s~\cite{CPP1,CPP2}, which marked the birth of 
particle physics both from an experimental and theoretical point of view. Soon after, Hincks and Pontecorvo~\cite{hincks} started to look
for the \meg~decay, and the missing observation of photons among the muon decay products demonstrated that it could not be considered 
just as an excited state of the electron. In the years 1950s the search for \meg~by Lokanathan and Steinberger~\cite{steinberger} and the exclusion 
of a branching ratio (BR) around $10^{-5}$ gave one of the first hints of the existence and conservation of lepton flavors, confirmed a few years
later by the discovery of the muon neutrino in the Nobel-awarded experiment by Lederman, Schwartz and Steinberger himself~\cite{muon-neutrino}.
Since then, studies of the muon decays have been carried on, till very recently~\cite{twist}, to test the structure of weak interaction.

The SM passed over the year all the experimental tests that have been performed, with only few (marginal or theoretically ambiguous) deviations 
of measurements from the predicted values (most notably, in the anomalous magnetic moment of the muon, $g_\mu-2$~\cite{g-2}). Nonetheless, 
the model looks incomplete from the theoretical point of view, due to some relevant limitations. Among them, a missing explanation
for some of the most striking features of the model (namely, the hierarchical structure of the particle mass spectra) and the indication
of a new mass scale around the TeV, that should prevent the mass of the Higgs boson to be pushed by radiative corrections to values much higher 
than the measured one. Many New Physics (NP) models have been built to overcome these difficulties and experiments are carried on to 
test these hypothetical scenarios.

Being the lightest unstable particle in the SM, with only one decay mode and its radiative derivates, and a relatively long lifetime (which
allow to produce intense beams even below the ultrarelativistic limit), the muon with its decays provides a unique environment to 
search for NP phenomena in a clean environment with essentially no theoretical uncertainty.

In this paper, the current status of the searches for physics beyond the SM in muon decays is reviewed. After a
short introduction to the experimental techniques used in this field (Sec.~\ref{sec:exp}), muon decays in the SM are discussed in
Sec.~\ref{sec:standard}. In Sec.~\ref{sec:lfv} the conservation of the lepton flavors is discussed, and the current status of the searches for
lepton-flavor-violating muon decays into SM particles is presented in Sec.~\ref{sec:lfv-exp}. Some more exotic scenarios with new
particles produced in muon decays are also discussed in Sec.~\ref{sec:exotics}.
 
\section{Experimental techniques for muon decay studies}
\label{sec:exp}

As already mentioned, the muon was discovered and firstly studied in the cosmic ray radiation. A significative step forward in the study of
its properties and decay modes was performed when pion beam became available in the years 1950s. Pions can be stopped in a
target and the muons produced in their decays can be studied. Finally, the first muon beams were delivered in the years 1970s.

Muon beams are usually produced starting from a proton beam impinging on a target and producing pions. Pions decaying in the 
target itself produce muons that can be collected and transported by a beam line toward the experimental areas where the detectors
are installed. The most intense continuous muon beams in the world~\cite{psi-beams,psi-target} are currently delivered at the 
Paul Scherrer Institut (PSI), in Villigen (CH). A 2~mA current of protons of 590~MeV kinetic energy from the PSI Ring Cyclotron hits 
two different graphite targets where, in total, $\sim 18\%$ of the protons are stopped, the rest being preserved to serve a 
downstream neutron spallation source. The two targets serve different beam lines. In order to get an intense, pure and monoenergetic 
muon beam, particles with momentum $p \sim 28$~MeV/c are selected, corresponding to muons emitted by pions decaying at rest. It happens when the
pion decays right on the surface of the production target (\emph{surface muons}). The beam can be further purified by selecting particles with a 
given velocity (and hence a given mass) by means of a Wien filter, a superposition of orthogonal electric (E) and magnetic (B) fields through 
which only particles with the desired velocity $v=E/B$ go straight, while others are removed by collimators. Beams with up to a few $10^8$ 
surface muons per second can be currently delivered at PSI.

In the Ring Cyclotron at PSI protons travel in bunches with a repetition rate of about 50~MHz. The pion lifetime ($\sim 26$~ns) 
is large enough to spoil the time structure of the beam and make a practically continuous muon beam. For some applications, pulsed
beams are instead necessary. For instance, the Delivery Ring at Fermilab (Batavia, USA) used for the Mu2e experiment~\cite{mu2e} provides protons in bunches separated 
in time by 1695~ns. They hit a target and produce muons that mostly arrive to the experimental apparatus within a few hundred ns. 
After about 700 ns and before the next bunch, only the muon decay products are practically visible in the detectors, with a very low
contamination from other particles (and specifically pions) emerging from the production target.

Muons decay predominantly into an electron~\footnote{We will generally use \emph{electron} to indicate either $e^+$ or $e^-$ emitted in muon decays,
unless the context implies that positive muons are used and hence positrons are always produced.} 
and a pair of neutrinos, $\mu^+ \to e^+ \overline \nu_\mu \nu_e$, with photons possibly
irradiated by one of the charged particles involved in the process. Hence, the only particles that can be reasonably detected in experiments
to study the muon decays with high statistics are electrons and photons. Most often the muons are stopped in a thin target in order
to exploit the advantages of the decay-at-rest kinematics. Hence, the particles to be detected also have a quite low momentum, in the range
of a few 10~MeV/c. All these features limit significantly the kind of detectors adopted in these studies. 

Tracking devices with a very low material budget, typically installed within a magnetic field to realize a magnetic spectrometer, are the usual choice for 
the electron reconstruction. The low material budget is necessary to reduce the multiple Coulomb scattering experienced by the electrons, which is very 
relevant at such low momenta and tends to spoil the original kinematics. For this reason, gaseous detectors like multi-wire proportional chambers (MWPC)
or drift chambers~\cite{sauli} have been mostly used in the last decades. In the MEG experiment~\cite{meg-det} the use of drift chambers with a
very light structure allowed to keep the total number of radiation lengths along the whole positron trajectory below $2 \times 10^{-3}$, i.e. less than 
one single layer of conventional silicon detectors. More recently, the development of very thin solid state detectors~\cite{mu3e-hvmaps} added a new 
interesting option for tracking at very low momentum. Fast organic scintillators can be added at the end of the electron trajectory to provide a precise time 
measurement.

The reconstruction of photons, when necessary, is typically performed with calorimeters. Inorganic scintillating crystals were mostly used in the past, while the MEG 
experiment firstly introduced the use of large volumes of liquid Xenon~\cite{meg-det}. An interesting alternative which have been
exploited in the past~\cite{mega} is the photon conversion into an $e^+e^-$ pair in the interaction with a thin layer of dense material, 
and the reconstruction of the two charged particles in a magnetic spectrometer to infer the photon energy and direction. The advantages and
disadvantages of this technique will be discussed in Sec.~\ref{sec:meg}.

\section{Muon decays in the Standard Model}
\label{sec:standard}

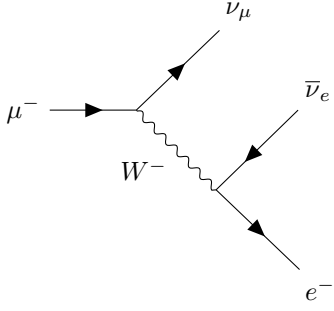
\begin{figure}[!t]
\centering
\begin{tikzpicture}
  \begin{feynman}
    \vertex (a) {\(\mu^{-}\)};
    \vertex [right=of a] (b);
    \vertex [above right=of b] (f1) {\(\nu_{\mu}\)};
    \vertex [below right=of b] (c);
    \vertex [above right=of c] (f2) {\(\overline \nu_{e}\)};
    \vertex [below right=of c] (f3) {\(e^{-}\)};
 
    \diagram* {
      (a) -- [fermion] (b) -- [fermion] (f1),
      (b) -- [boson, edge label'=\(W^{-}\)] (c),
      (c) -- [anti fermion] (f2),
      (c) -- [fermion] (f3),
    };
  \end{feynman}
\end{tikzpicture}
\caption{\label{fig:muon_decay}Feynman diagram of the Michel muon decay $\mu^+ \to e^+ \overline \nu_\mu \nu_e$.}
\end{figure}

In the SM the muon decays through weak interactions mediated by the $W$ boson, according to the Feynman diagram shown in Figure~\ref{fig:muon_decay}.
The lifetime, at tree level and neglecting the electron mass, can be written as~\footnote{We use here the system of units where $\hbar = c = 1$.}:
\begin{equation}
\tau_\mu = \left( \frac{m_\mu^5 G_F^2}{192\pi^3} \right)^{-1} \sim 2.2 \times 10^{-6}~\mathrm{s} \; ,
\end{equation}
where $G_F \sim 1.17~\mathrm{GeV}^{-2}$ is the Fermi coupling constant and $m_\mu \sim 105.7~\mathrm{MeV}/c^2$ 
is the muon mass. This result comes from the lagrangian terms:
\begin{eqnarray}
\mathcal{L}_{W\mu} &=& -\frac{g}{\sqrt{2}} \, \overline \nu_\mu \gamma^\mu (1 - \gamma^5) \mu \, W^+_\mu \; ,\\
\mathcal{L}_{We} &=& -\frac{g}{\sqrt{2}} \, \overline \nu_e \gamma^\mu (1 - \gamma^5) e \, W^+_\mu \; ,
\end{eqnarray}
that couple charged leptons and neutrinos to the $W$ boson, with $G_F/\sqrt{2} = g^2/(8m_W^2)$. Here $g$ is the coupling constant of the
weak interaction, and $m_W$ is the mass of the $W$ boson which mediates the decay in question. The $(1 - \gamma^5)$ term selects only the
left-handed components of the fermion fields and right-handed components of the anti-fermion fields, and
give rise to the maximally parity violating $V-A$ structure of weak interactions. It is reflected in the differential rate of polarized muon decays 
in the SM~\cite{kuno-okada}:
\begin{eqnarray}
\frac{d^2\Gamma(\mu^\pm \to e^\pm \overline \nu \nu)}{dx \, d\cos\theta_e} = \frac{m_\mu^5 G_F^2}{192\pi^3}x^2\left[ \left( 3-2x\right) \pm P_\mu\cos\theta_e(2x-1)\right] \; ,
\end{eqnarray}
where $x = 2E_e/m_\mu$, $E_e$ is the electron energy and $\theta_e$ is the angle of the electron with respect to the polarization axis of a muon 
beam with polarization degree $P_\mu$.

\begin{figure}[!t]
\centering
\includegraphics[scale=.4]{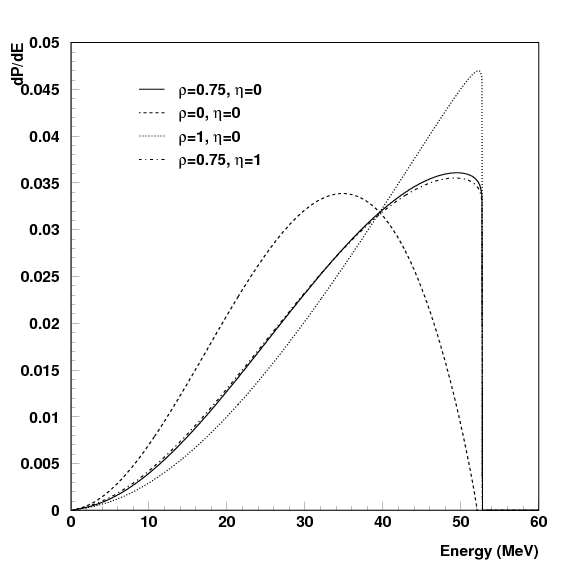}
\caption{\label{fig:michel}Spectrum of the electron energies from the $\mu^+ \to e^+ \overline \nu_\mu \nu_e$ decay, in the SM and in case of deviations
of the $\rho$ and $\eta$ parameters from the SM values (picture taken from~\cite{icarus}).}
\end{figure}

A more general form, including non-standard interactions and summed over the electron polarizations, is~\cite{fetscher}:
\begin{eqnarray}
\frac{d^2\Gamma(\mu^\pm \to e^\pm \overline \nu \nu)}{dx \, d\cos\theta_e} = \frac{m_\mu G_F^2}{4\pi^3} W_{e\mu}^4 \sqrt{x^2-x_0^2} \left[ F_{IS}(x) \pm P_\mu\cos\theta_e F_{AS}(x)\right] \; ,
\end{eqnarray}	
where $W_{e\mu} = (m_\mu^2 + m_e^2)/(2m_\mu)$, $x_0 = m_e/W_{e\mu}$ and:
\begin{eqnarray}
F_{IS}(x) &=& x(1-x) + \frac{2}{9}\rho(4x^2 - 3x - x_0^2) + \eta x_0(1-x) \; ,\\
F_{AS}(x) &=& \frac{1}{3}\xi \sqrt{x^2 - x_0^2}\left\{ 1 - x + \frac{2}{3}\delta\left[4x - 3 + \left( \sqrt{1 - x_0^2} - 1\right)\right] \right\} \; .
\end{eqnarray}
The parameters $\rho$, $\eta$, $\xi$ and $\delta$ are called Michel parameters~\cite{michel} and their SM values are $\rho = 3/4$, $\eta = 0$, $\xi = 1$
and $\delta = 3/4$. The $\mu^+ \to e^+ \overline \nu_\mu \nu_e$ decay is often referred to as the Michel decay. Precise measurements of these parameters
allow to verify the V-A structure of weak interactions and search for possible contributions beyond the SM. The energy spectrum 
reaches its maximum and has a sharp edge around the kinematical end-point of the electron energy, $E_e \sim 52.8$~MeV, 
as shown in Figure~\ref{fig:michel}.

A photon with energy $E_\gamma > 10$~MeV can be emitted by one of the charge particles involved in the process~\cite{rmd}, with a probability of about 1.4\%,
producing the radiative muon decay (RMD) $\mu^+ \to e^+ \overline \nu_\mu \nu_e \gamma$. The photon emitted in this process can be a virtual photon 
converting into an $e^+ e^-$ pair. The resulting $\mu^+ \to e^+ \overline \nu_\mu \nu_e e^+ e^-$ decay has a BR of $3.4 \times 10^{-5}$~\cite{mu3enunu}.

\subsection{Searches for New Physics in standard muon decays}

\begin{figure}[!t]
\centering
\includegraphics[scale=.7]{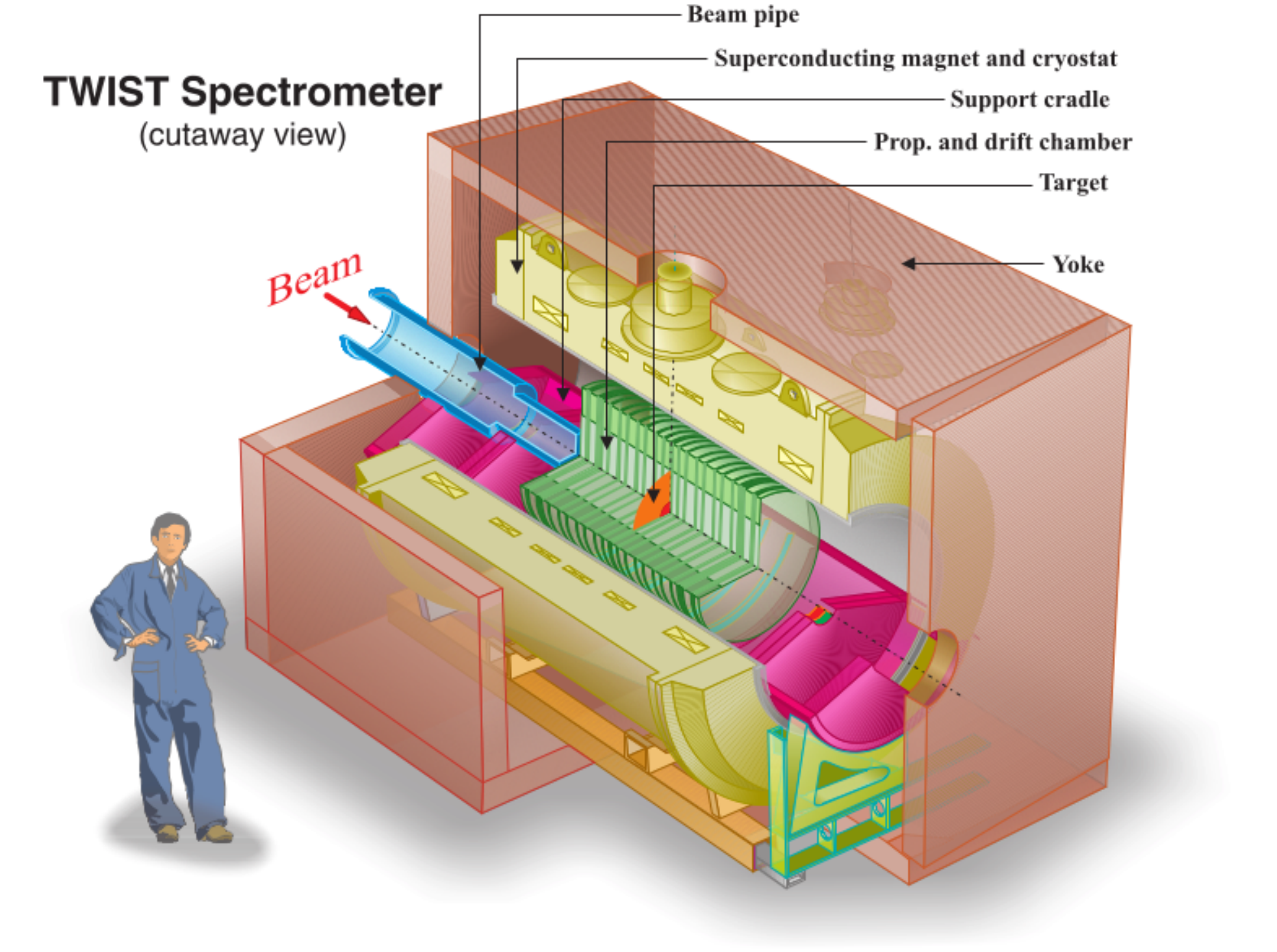}
\caption{\label{fig:twist}A sketch of the TWIST experiment.}
\end{figure}

As already mentioned, the study of the spectrum of the electrons emitted in the Michel decay allow for accurate tests of the SM, with negligible
theoretical uncertainties (see the Muon Decay Parameters review in~\cite{pdg} for a detailed discussion). The latest results in this field come from 
the TWIST experiment~\cite{twist}, operated at TRIUMF (Vancouver, CA). 
Positive muons are used in this experiment, because negative muons tend to be captured by the Coulomb field of the target nuclei, and the
decay spectra are distorted with respect to free muons. A muon stopping rate between 2000 and 5000~s$^{-1}$ was available, and 
a total of $11 \times 10^{9}$ events were collected, although only a portion of the observed spectrum was retained for the final analysis,
concentrating on the phase space regions where minimal discrepancies between data and simulations were present, 
in order to reduce the associated systematic uncertainties. Including all the selection criteria, $0.55 \times 10^9$ events were 
in fact used. The TWIST detector was composed of proportional chambers and drift chambers surrounding a silver or aluminum 
muon stopping target, minimally degrading the muon polarization, and immersed in a 2~T solenoid field. 
A sketch of the experiment is shown in Figure~\ref{fig:twist}.
A very low reconstruction inefficiency ($< 10^{-4}$) was obtained over the measured phase space region, with a momentum
resolution of 58~keV/c at $\sin\theta_e = 1$ and a momentum accuracy $O(10^{-4})$. These excellent performances allowed to measure the Michel
parameters with high precision~\cite{twist-parameters,twist-weak}:
\begin{eqnarray}
\rho &=& 0.74977 \pm 0.00012~\mathrm{(stat)} \pm 0.00023~\mathrm{(syst)} \; ,\\
\delta &=& 0.75049 \pm 0.00021~\mathrm{(stat)} \pm 0.00027~\mathrm{(syst)} \; ,\\
P_\mu^\pi &=& 1.00084 \pm 0.00029~\mathrm{(stat)} ^{+0.00165}_{-0.00063}~\mathrm{(syst)} \; ,
\end{eqnarray} 
where $P_\mu^\pi$ is the muon polarization in pion decays (equal to 1 in the SM). No significative deviation with respect to the SM values is observed.

A precise measurement of $\eta$ requires a measurement of the electron polarization~\cite{eta}. It can be made by means of a magnetized foil 
used as a polarization analyzer: in such a foil, positrons annihilate in flight with electrons and the angular distribution of the photons emitted in this process 
is related to the polarization of the positrons. The best estimate reported in~\cite{pdg} is $\eta = 0.057 \pm 0.034$.

Contributions beyond the SM can be also searched for in RMDs~\cite{bogart,vdShaaf,pibeta}. In this case, deviations from the V-A structure of weak interactions 
should be observed in the electron and photon spectra and described by a parameter $\overline \eta$ that is zero in the SM~\cite{lenard,behrends,fronsdal}. 
The best measurement~\cite{pibeta} is currently $\overline \eta = 0.006 \pm 0.017~\mathrm{(stat)} \pm 0.018~\mathrm{(syst)}$.

The MEG experiment, which will be discussed in detail in Sec.~\ref{sec:meg}, also performed a measurement of RMD spectra~\cite{meg-rmd}, with the main goal
of cross-checking the background predictions for the search of \meg. As a by product, it allowed to verify the SM prediction in a limited region of the
phase space (large momentum and large $e\gamma$ relative angle), but with very high statistics.

\section{A lepton flavor violation primer}
\label{sec:lfv}

In the SM fermions are divided in leptons, which only feel the electroweak interactions, and quarks, which are also subject to the strong interaction. Quarks
are divided in three families:
\begin{equation}
\label{eq:weak-states}
\left( \begin{array}{c} u \\ d \end{array} \right) \; , \; \left( \begin{array}{c} s \\ c \end{array} \right) \; , \; \left( \begin{array}{c} t \\ b \end{array} \right) \; ,
\end{equation}
and weak interactions mediated by the $W^{\pm}$ boson transform at tree level the upper components into the lower components of these doublets, according
to the lagrangian term:
\begin{equation}
\label{eq:weak}
\mathcal{L}_{Wq} = -\frac{g}{\sqrt{2}} \left[ \overline d \gamma^\mu (1 - \gamma^5) u \, W^+_\mu  + \overline c \gamma^\mu (1 - \gamma^5) s \, W^+_\mu + \overline b \gamma^\mu (1 - \gamma^5) t \, W^+_\mu + h.c. \right] \equiv -\frac{g}{\sqrt{2}}\left[ \mathbf{\overline d} \gamma^\mu (1 - \gamma^5) \mathbf{u} \, W^+_\mu \right]\; ,
\end{equation}
with:
\begin{equation}
\mathbf{u} = \left( \begin{array}{c} u \\ s \\ t \end{array} \right) \; , \; \mathbf{d} = \left( \begin{array}{c} d \\ c \\ b \end{array} \right) \; .
\end{equation}
The quark states of Eq.~(\ref{eq:weak-states}) are the eigenstates of the weak interaction hamiltonian, but they do not coincide necessarily with the
mass eigenstates (i.e. the eigenstates of the free Hamiltonian), which will be in general a linear combination of the former, $\mathbf{u'} = U_u \mathbf{u}$ and 
$\mathbf{d'} = U_d \mathbf{d}$. If the lagrangian of Eq.~(\ref{eq:weak}) is rewritten in terms of the mass eigenstates, then:
\begin{equation}
\mathcal{L}_{Wq} = -\frac{g}{\sqrt{2}}\left[ \mathbf{\overline d'} \gamma^\mu (1 - \gamma^5) V \mathbf{u'} \, W^+_\mu \right] \; ,
\end{equation}
where $V = U^\dagger_d U_u$ is the Cabibbo-Kobayashi-Maskawa (CKM) matrix~\cite{cabibbo,KM}. Hence, the weak interaction couples the up-like mass eigenstates to 
linear combinations of the down-like mass eigenstates, and they are said to \emph{mix} the families.

In the lepton sector, we also consider three families (\emph{flavors}):
\begin{equation}
\label{eq:lepton-states}
\left( \begin{array}{c} e^- \\ \nu_e \end{array} \right) \; , \; \left( \begin{array}{c} \mu^- \\ \nu_\mu \end{array} \right) \; , \; \left( \begin{array}{c} \tau^- \\ \nu_\tau \end{array} \right) \; ,
\end{equation}
and, with similar notations, the lagrangian:
\begin{equation}
\mathcal{L}_{W\ell} = -\frac{g}{\sqrt{2}}\left[ \mathbf{\overline N} \gamma^\mu (1 - \gamma^5) \mathbf{L} \, W^+_\mu \right]\; ,
\end{equation}
In this case, if we consider the basic version of the SM where the neutrinos are massless, and also consider that neutrinos are only subject to weak
interactions, we can arbitrarily redefine the neutrino weak eigenstates in such a way that the lagrangian is still diagonal in the charged-lepton mass eigenstates. 
It implies that the weak interactions do not mix the lepton flavors and, since electromagnetic and strong interactions cannot mix the flavors, the number of leptons 
minus the number of anti-leptons of each family is separately conserved in any process (\emph{lepton flavor conservation}).

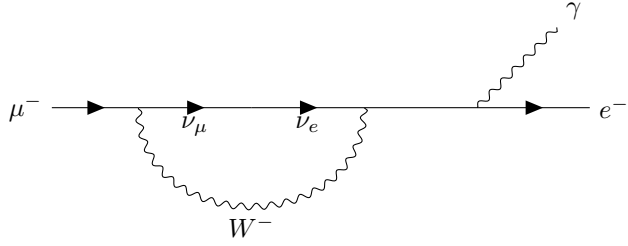
\begin{figure}[!t]
\centering
\begin{tikzpicture}
  \begin{feynman}
    \vertex (a) {\(\mu^{-}\)};
    \vertex [right=of a] (b);
    \vertex [right=of b] (c);
    \vertex [right=of c] (d);
    \vertex [right=of d] (e);
    \vertex [right=of e] (f) {\(e^{-}\)};
    \vertex [above right=of e] (g) {\(\gamma\)};
 
    \diagram* {
      (a) -- [fermion] (b) -- [fermion, edge label'=\(\nu_{\mu}\)] (c) [dot],
      (c) -- [fermion, edge label'=\(\nu_{e}\)] (d) -- (e) -- [fermion] (f),
      (b) -- [boson,half right, edge label'=\(W^{-}\)] (d),
      (e) -- [boson] (g),
    };
  \end{feynman}
\end{tikzpicture}
\caption{\label{fig:lfv-SM}Feynman diagram of the $\mu \to e \gamma$ decay in the SM. The photon can be emitted by any charged particle in the process.}
\end{figure}

In reality, this lepton flavor symmetry is only approximate in the SM. Oscillations from a neutrino flavor to another have been indeed 
observed~\cite{kamiokande,SNO}, and they imply non-zero neutrino masses, so that the arbitrarily redefinition of the fields discussed above 
is not possible and flavor mixing arises. It is parameterized by the lepton-sector analog of the CKM matrix, the Pontecorvo-Maki-Nakagawa-Sakata (PMNS)
matrix~\cite{pontecorvo,mns}. The probability that a neutrino of energy $E$ and flavor $\alpha$ oscillates into a neutrino of flavor $\beta$ over a distance $L$ is
given by:
\begin{equation}
P(\nu_\alpha \to \nu_\beta) = \left| \sum_i U^*_{\alpha i} U _{\beta i} e^{-i m_{\nu_i}^2 L/2E}  \right|^2 \; ,
\end{equation}
where $U$ is the PMNS matrix and $m_{\nu_i}$ are the neutrino masses, not yet measured and constrained to be in the sub-eV range by the
present experimental results. Flavor mixing also arises in the charged-lepton sector, where for instance the \meg~decay 
depicted in Figure~\ref{fig:lfv-SM} can be conceived as the oscillation of a neutrino over a length $O(1/m_W)$ with a characteristic energy $O(m_W)$.
The process is thus suppressed by the tiny factors $O(m_{\nu_i}^4/m_W^4)$. The exact calculation of the BR of the \meg~decay gives:
\begin{equation}
BR(\meg) = \frac{3\alpha}{32\pi} \left| \sum_{i=1,3} U^*_{\mu i} U_{e i} \frac{m_{\nu_k}^2}{m_W^2}\right|^2
\end{equation}
and, with the best estimates of the PMNS elements, $\mathrm{BR}(\meg) \sim 10^{-54}$~\cite{calibbi-signorelli}. Compared to the current best experimental
upper limit (UL), $\mathrm{BR}(\meg) < 4.2 \times 10^{-13}$ at 90\% confidence level (CL), the expected SM rate is unmeasurably small. Similar
considerations hold for the other charged lepton flavor violation (cLFV) processes in the SM.

From the discussion above it is clear that, on one side, the observation of cLFV would be an unambiguous evidence of NP beyond the SM, with no
theoretical uncertainty. On the other side, the lepton flavor symmetry is accidental: it is not related to the structure of the SM lagrangian, but it is 
only a mere consequence of the particle content of the model. This feature makes this symmetry prone to be violated in any extension of the
SM. Indeed, most NP models predict an enhancement of cLFV, and they are already strongly constrained by the present experimental limits.
Supersymmetric (SUSY) models are the most popular choice, and searches for cLFV tends to give strong constraints to them, owing to an interesting feature:
even if a SUSY model is built to be lepton-flavor-conserving at some high energy scale, the renormalization group evolution creates
unavoidable off-diagonal terms in the slepton mass matrix, producing cLFV effects at a level that is typically measurable in present or next generation experiments
(a discussion of cLFV in SUSY, with a reach bibliography, can be found in~\cite{calibbi-signorelli}). It is worth mentioning, anyway, that cLFV effects
of the same size typically arise in a plethora of NP models, including theories with extra dimensions~\cite{ED}, multi-Higgs models~\cite{multi-higgs},
unparticle physics~\cite{unparticles}, leptoquarks~\cite{leptoquarks}. The importance of cLFV searches on the development of NP models can be easily 
understood in the framework of effective field theories (EFT), where non-standard operators are included in the SM lagrangian, suppressed by powers of a 
characteristic mass scale $\Lambda$. If one assumes coefficients $O(1)$ in front of these new contributions and takes into account the current limits on cLFV, 
NP is excluded up to thousands of TeV~\cite{deGouvea}. So we already know that, if we want some physics beyond the SM at the TeV scale, some protection mechanism 
is likely to exist, that keep the coefficients low enough to satisfy the current bounds.
 
\begin{figure}[!t]
\centering
\includegraphics[scale=.5]{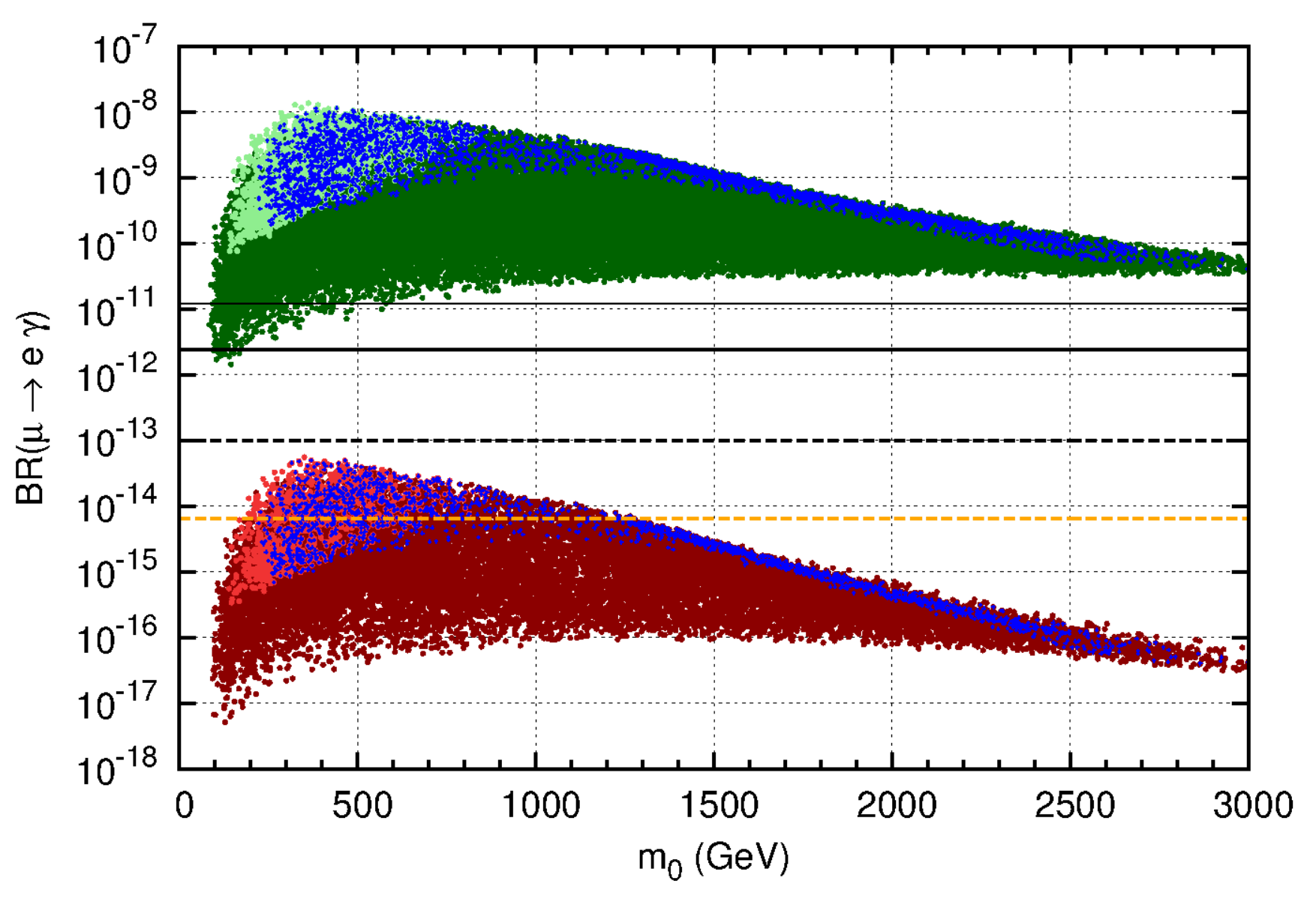}
\caption{\label{fig:susy}Predictions for \meg~in a specific Constrained Minimal Supersymmetric Standard Model as a function of the universal 
scalar mass $m_0$~\cite{susy}. Dots correspond to points of the SUSY parameter space that are accessible at the LHC. Red points are obtained in a 
scenario where the slepton mixing matrix is CKM-like (minimal mixing), while for green points it is PMNS-like (maximal mixing). The full black line 
shows the MEG limit in 2011~\cite{meg-2011}.}
\end{figure}

Once such a protection is implemented, it is still interesting to consider the interplay between indirect searches of NP through low-energy cLFV 
measurements and direct searches of new particles at the high-energy frontier with accelerators like the LHC. In Figure~\ref{fig:susy} the cLFV predictions of
a specific SUSY model~\cite{susy} are illustrated. Points in the plot correspond to points of the model's parameter space that would give accessible
signatures at the LHC, and different scenarios are assumed. It is noticeable that, even within the same model, there are cases where 
possible LHC signatures are already completely excluded by cLFV searches, and others where cLFV searches can give almost no clue about the LHC reach. 
This is a typical feature in most models and it makes clear the complementarity between the two kind of searches: both are necessary to increase the 
chance of making a discovery and, if one is made, both are needed to disclose the flavor structure of the observed NP.

\section{Lepton flavor violating muon decays to standard particles}
\label{sec:lfv-exp}

In this section we discuss the most recent and future searches for cLFV in muon decays, with standard particles in the final state, while the next section
will be devoted to some more exotic scenarios where new particles are also produced.

\begin{figure}[!t]
\centering
\hspace{-4cm}
\includegraphics[scale=.5]{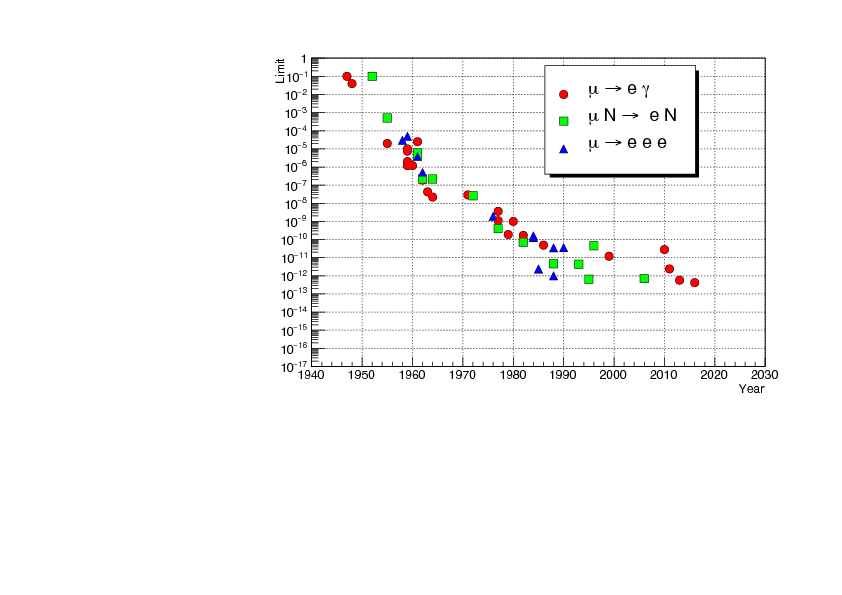}
\vspace{-3cm}
\caption{\label{fig:history}Historical evolution of the limits on \meg~(Sec.~\ref{sec:meg}), \meee~(Sec.~\ref{sec:meee}) and 
$\mu \to e$ conversion in nuclei (Sec.~\ref{sec:mu2e}).}
\end{figure}

The historical evolution of the searches for LFV muon decays is shown in Figure~\ref{fig:history}. It should be noticed anyway that, as we will discuss in 
this section, the impact of these searches in terms of constraints to NP effects cannot be trivially evaluated 
from the absolute values shown in this picture.

In the past, searches have been performed also for $\mu^- \to e^- \nu_e \overline \nu_\mu$, and the best upper limit on the BR of this decay is 1.2\%~\cite{freedman}.
In most NP scenarios this kind sensitivity is far from being competitive with the one of the other experiments, and these experiments will be not discussed 
here, although they contribute to constrain some models~\cite{bileptons}.

\subsection{\meg}
\label{sec:meg}

The \meg~decay has been searched for since the discovery of the muon. As already mentioned, the first search was performed in 1948 by 
Hincks and Pontecorvo~\cite{hincks}, who searched for photons produced in association with the electron typically observed in the decay of cosmic muons. 
Later, the search was performed using stopped pion beams and finally with muon beams. In this case, surface muons ($\sim 28$~MeV/c)
are preferred to guarantee high rate and beam purity. Muons are stopped on a thin target (typically, a few hundred microns of plastic) and the 
kinematics of the two-body decay at rest is exploited, searching for a positron and a photon with the same energy 
$E_e = E_\gamma = m_\mu/2 \sim 52.8$~MeV (we neglect here the positron mass), emitted in opposite directions 
(i.e. with a relative angle $\Theta_{e\gamma} = 180^\circ$). Notice that positive muons are used, because negative muons are captured by the target nuclei
and the resulting decay spectra are distorted with respect to the free muon case.

There is a physics background for this decay channel, given by the RMD $\mu^+ \to e^+ \overline \nu_\mu \nu_e$, when the two neutrinos carry a very
low fraction of the available energy, so that the photon and positron energies and relative angle are compatible with the two-body kinematics within the 
experimental resolutions. The energy and angle spectra near the kinematical end point are anyway suppressed by terms proportional to 
$(1 - 2E_e/m_\mu)^l\,(1 - 2E_\gamma/m_\mu)^m\,(\pi-\Theta_{e\gamma})^n$, with $l+m+n = 6$, as shown in~\cite{kuno-okada}. This strong suppression,
combined with a very high muon beam rate $\Gamma_\mu$ (exceeding $10^7~\mu$/s in the most recent experiments), makes the RMD background secondary with
respect to the accidental time coincidence of a positron from a muon decay and a photon from the RMD of another muon or the annihilation in flight (AIF)
of another high energy positron (the AIF contribution being dominant near the kinematical end point). In this case, anyway, a further handle to suppress the 
background comes from the requirement that the two particles are emitted at the same time. There are hence four discriminating variables to be exploited to 
separate signal and background: $E_e$, $E_\gamma$, $\Theta_{e\gamma}$ and the relative time $T_{e\gamma}$. If one imagines to extract the signal 
yield by counting events in this 4-dimensional space, within a signal region defined according to the resolutions $\delta E_e$, $\delta E_\gamma$, etc., 
the expected accidental background rate is~\cite{kuno-okada}:
\begin{equation}
\label{eq:acc}
\Gamma_\mathrm{acc} = \Gamma_\mu^2 \, \delta E_e \, (\delta E_\gamma)^2 \, (\delta\Theta_{e\gamma})^2 \, \delta T_{e\gamma} \; ,
\end{equation}
where the $\delta E_e$ and $(\delta E_\gamma)^2$ terms follow from the shape of the Michel and AIF spectra, while the $(\delta\Theta_{e\gamma})^2$ factor
can be understood if it is decomposed in two components in spherical coordinates $\delta\phi_{e\gamma}\delta\theta_{e\gamma}$. The first conclusion we
draw from Eq.~(\ref{eq:acc}) is that, if an experiment with a given $\Gamma_\mu$ expects to see over its lifetime a significative number of accidental events in 
its signal region, an increase of the beam rate would be useless to improve the sensitivity, that would go with 
$\Gamma_\mu/\sqrt{\Gamma_\mathrm{acc}}$ and would remain constant. Hence, an experiment designed to search for \meg~needs to provide good angle, 
time and positron energy resolutions, and an excellent photon energy resolution, and only if these performances are good enough to guarantee a zero-background 
condition the muon beam rate can be profitably increased.

A good reconstruction of the positron direction is typically guaranteed by very light tracking detectors, and gaseous detectors have been used in the 
most recent experiments. If the detector is installed within a magnetic field, the positron momentum is also measured. The requirement of using 
light detectors allows to reduce the deterioration of the resolutions due to the multiple Coulomb scattering, and also reduces the rate of AIF events contributing
to the accidental background. For the same reason, the stopping target also needs to be as light as possible, while keeping a high muon stopping efficiency.
The strategy adopted to fulfill this requirement consists in the use of a thin planar stopping target, with its normal vector forming a large angle with respect 
to the beam axis. In this way, positrons exiting around $90^\circ$ with respect to the beam axis see a much lower thickness of target material with respect 
to the muons. A thin planar target also allows to define the muon decay point by simply propagating the positron track back to the target plane. Combined
with a measurement of the photon detection points, it gives for signal events the best estimate of the photon direction (as the line joining the two points), 
circumventing the typically bad (or even null) angular resolution of $\gamma$ detectors. Fast organic scintillators are also used at the end of the positron trajectory
to provide a precise timing.

On the photon side, a calorimetric approach with inorganic scintillators guarantees a precise measurement of energy, position and time, typically limited by the
properties (light yield and characteristic time) of the scintillating material. The efficiency is mostly limited by the material in front of the calorimeter, because the
photon can convert before entering it. It can be relevant, for instance, when the coils of the magnet of a positron spectrometer are present. 
In the MEG experiment~\cite{meg-det}, for instance, it determined an inefficiency of about 35\%. There is indeed an alternative approach for the photon detection, 
that consists in placing within the magnetic field a thin layer of dense material where the photon can convert into an $e^+e^-$ pair, which can be tracked in the 
magnetic field in order to infer the photon energy, conversion position and direction. In this case the efficiency is very low, due to
the low probability of photon conversion (4.5\% in $\sim 0.1$ radiation lengths of Lead or Tungsten~\cite{meg-future} for 52.8~MeV photons). On the
other hand, the reconstruction of the $e^+e^-$ pair provides: an excellent energy resolution; an excellent photon position resolution; an information
on the photon direction which is not competitive with the resolution provided by the line joining the photon position and the positron vertex, but can 
be compared to this one in order to determine if the photon really comes from the same point of the positron (which helps to reduce the accidental background).
According to the discussion above about the accidental background rate, the resolutions and background rejection of the photon conversion approach
allow to increase the beam rate, and it can compensate the loss of efficiency.

The most recent experiments looking for \meg~made use of all these techniques.

\begin{figure}[!t]
\centering
\includegraphics[scale=.5]{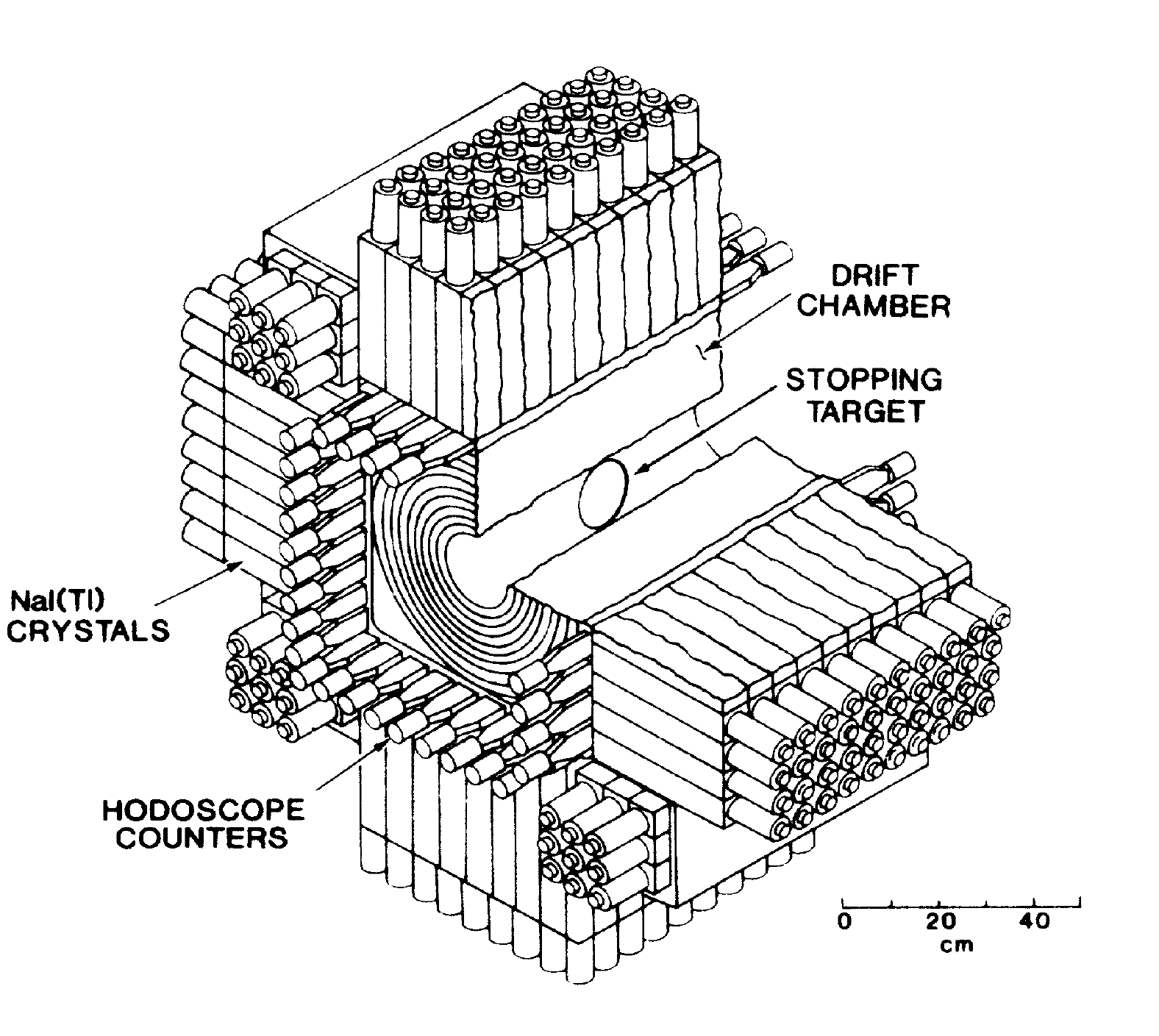}
\includegraphics[scale=1.5]{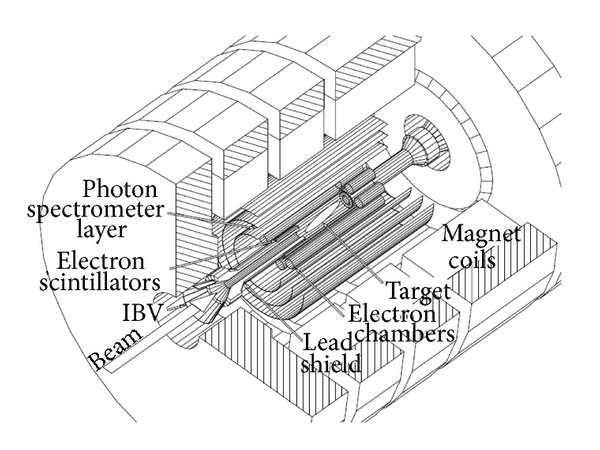}
\caption{\label{fig:cb-mega}A sketch of the Crystal Box (left) and MEGA (right) experiments.}
\end{figure}

The Crystal Box experiment~\cite{CrystalBox} operated at LAMPF (Los Alamos, USA) in the years 1980s and depicted 
in Figure~\ref{fig:cb-mega}, was composed of a multilayer cylindrical drift chamber surrounded by scintillator counters for positron timing 
and a calorimeter made of NaI(Tl) crystals. There was no magnetic field, 
hence the drift chamber only provided the positron direction, while energy measurements were performed in the calorimeter for both the positron and the electron.
The experiment reached $E_e, E_\gamma$ resolutions $< 8\%$ (FWHM) at 52.8~MeV, a $T_{e\gamma}$ resolution of 1.15~ns (FWHM)
dominated by the photon timing in the calorimeter, and a vertex resolution of a few mm. The photon reconstruction efficiency was exceeding 98\%
and the total efficiency for \meg~events was 64\%. Running at $4 \times 10^{5}~\mu$/s of beam intensity, the experiment collected a statistics 
corresponding to $1.4 \times 10^{12}$ muons stopped on target. In the parameter space defined by the four discriminating variables 
$E_e$, $E_\gamma$, $\Theta_{e\gamma}$ and $T_{e\gamma}$, in a region around the expected signal values, data were analyzed with a 
Maximum Likelihood technique (this approach was also used in later experiments). No significative signal was observed and an UL was set, 
$\mathrm{BR}(\meg) < 4.9 \times 10^{-11}$ at 90\% CL. Crystal Box also searched for $\mu^+ \to e^+ \gamma \gamma$ and set the limit 
$\mathrm{BR}(\mu^+ \to e^+ \gamma \gamma) < 7.2 \times 10^{-11}$.

The MEGA experiment~\cite{mega-prd}, also run at LAMPF about 10 years later, could exploit a much larger beam rate, up to $4 \times 10^7~\mu$/s.
With such a rate, it was reasonable to exploit the better resolutions provided by the photon conversion technique. The experiment, sketched in Figure~\ref{fig:cb-mega}, was composed of 
a positron spectrometer with MWPCs within a 1.5~T magnetic field, surrounding a $0.1$~mm Mylar target whose normal direction formed an angle of $83^\circ$ 
with respect to the beam axis. Around the positron spectrometer and within the same field, two layers of Lead, 0.045 radiation lengths each, were used for the 
photon conversion. MWPC in between the two layers and a drift chamber after the second layer tracked the $e^+e^-$ pair. The photon reconstruction 
efficiency was only 2.4\%, but an energy resolution of 5.7\% and 3.3\% (FWHM), for conversion in the inner and outer layer respectively, could be reached. 
The positron spectrometer provided a momentum resolution of 540~keV/c. A time resolution of 2~ns was provided by scintillators in between the positron 
MWPC and the conversion layers. The experiment collected data for three years for a total of $8 \times 10^6$~s of live time and $1.2 \times 10^{14}$ muons 
stopped on target. No excess of events was observed and the limit $\mathrm{BR}(\meg) < 1.2 \times 10^{-11}$ at 90\% CL was set~\cite{mega}. 

\begin{figure}[!t]
\centering
\includegraphics[scale=.2]{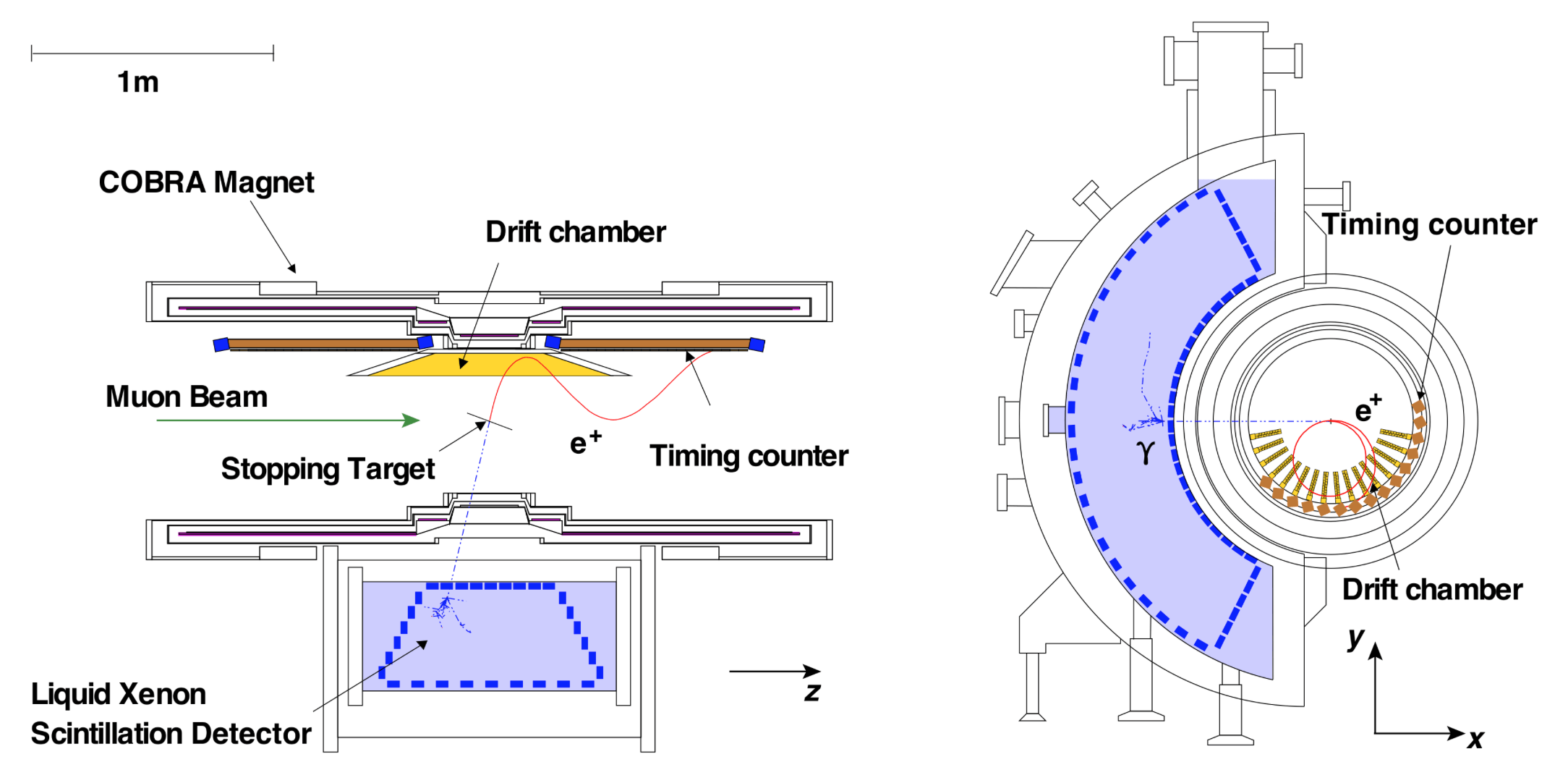}
\caption{\label{fig:meg}A sketch of the MEG experiment.}
\end{figure}

\begin{figure}[!t]
\centering
\includegraphics[scale=.8]{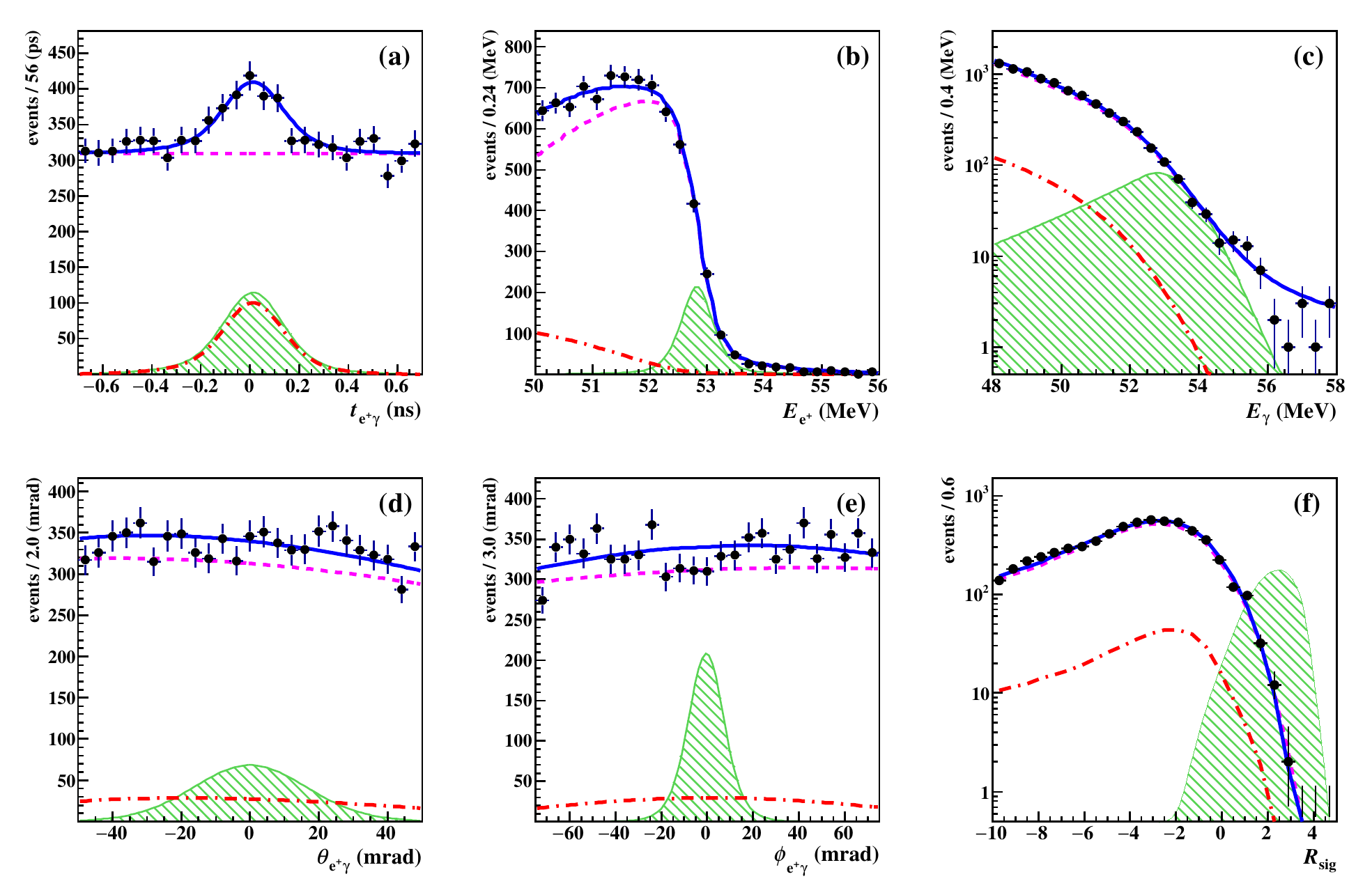}
\caption{\label{fig:meg-ana}Distributions of the five discriminating variables $\mathbf{x} = (T_{e\gamma},E_e,E_\gamma,\theta_{e\gamma},\phi_{e\gamma})$
used in the Likelihood analysis of data from the MEG experiment. 
The bottom-right plot shows the distribution of the variable $R_{sig} = \log_{10}(S(\mathbf{x})/(f_R R(\mathbf{x}) + f_A A(\mathbf{x}))$, where 
$S$, $R$ and $A$ are the multidimensional PDFs for signal, RMD and accidental background, while $f_R$ and $f_A$ are the expected fractions
of RMD and accidentals. This quantity summarizes the discriminating power of the likelihood analysis but is not used in the extraction of the limit.
Dots: experimental data. Magenta dashed line: accidental background PDF. Red dot-dashed line: RMD background PDF. 
Green hatched area: signal PDF assuming $BR(\meg) = 4.2 \times 10^{-11}$ (100 times the UL).}
\end{figure}

The best \meg~limit currently available comes from MEG~\cite{meg-det}. The main breakthrough of this experiment was the construction of
a $900~\ell$ liquid Xenon (LXe) calorimeter read out by 846 PMTs, providing excellent time (64~ps RMS) and energy (1.7 to 2.4\% RMS) 
resolutions, even better than what could be obtained in MEGA with the photon conversion approach, although the extremely high cost of LXe limited the volume of the 
detector and hence its acceptance ($\sim 10\%$). The experiment is depicted in Figure~\ref{fig:meg}. For positron tracking, a set of 16 planar drift chambers 
was immersed in a magnetic field which was degrading from 1.3~T at the center of the detector to about 0.5~T at the extremities of the magnet, with
its main component along the beam axis. This configuration has two advantages: positrons of same momentum make a trajectory with the same bending 
radius, irrespective of the emission angle, and even tracks emitted at almost $90^\circ$ with respect to the beam axis are expelled from the spectrometer 
after a few turns. It made possible to keep a high acceptance in the chambers, even if they only cover a small extent in the radial direction, which allows to 
keep a low pileup of low momentum positrons. The momentum resolution was $\sim 330$~keV/c and the angular resolutions around 10~mrad.
At the end of their trajectory the positrons reach a set of scintillating bars for positron timing (with 65~ps RMS resolution) and trigger.
The experiment run at PSI with $3 \times 10^7~\mu$/s, collecting up to $7.5 \times 10^{14}$ muons stopped on target. A likelihood analysis was performed,
splitting $\Theta_{e\gamma}$ in two components in spherical coordinates, $\theta_{e\gamma}$ and $\phi_{e\gamma}$. The distributions of data
are shown in Figure~\ref{fig:meg-ana} along with the probability distribution functions (PDF) used in the likelihood analysis. No excess of signal events was 
observed and an UL at 90\% CL was set, $\mathrm{BR}(\meg) < 4.2 \times 10^{-13}$~\cite{meg}.

The MEG collaboration is finalizing the construction of an upgraded detector, MEG-II~\cite{meg-upgrade}, with significative improvements in all its components. 
The 16 drift chambers of MEG have been replaced by a unique cylindrical drift chamber, which will provide a better tracking along the full trajectory of the
positron, and will reduce the inefficiencies associated in MEG to the long untracked path between the chambers and the positron scintillators. The expected
momentum resolution of the spectrometer is expected to be $\sim 130$~keV/c while the angular resolutions will be improved by a factor of 2 with respect to MEG.
The LXe calorimeter has been instrumented in its inner face with silicon photon detectors (MPPC)~\cite{mppc}, giving a higher granularity with respect to PMTs,
and hence improving the energy and position resolution of the detector. The energy resolution is expected to be $\sim 1\%$. The positron scintillating bars 
are replaced by scintillating tiles, in such a way that many time measurements are performed along the positron trajectory. The average positron time resolution
is expected to go below 40~ps. Finally, a new detector has been installed along the beam line in order to identify a possible low energy positron in coincidence with
the photon in the LXe calorimeter, which would indicate that the photon itself comes from a RMD and not from a signal event. All the detectors have been installed 
at the end of 2018 and data taking will start in 2019, for a three year run at $7.7 \times 10^7~\mu$/s and an expected UL of $6 \times 10^{-14}$ on the 
BR of \meg.

\begin{figure}[!t]
\centering
\includegraphics[scale=.4]{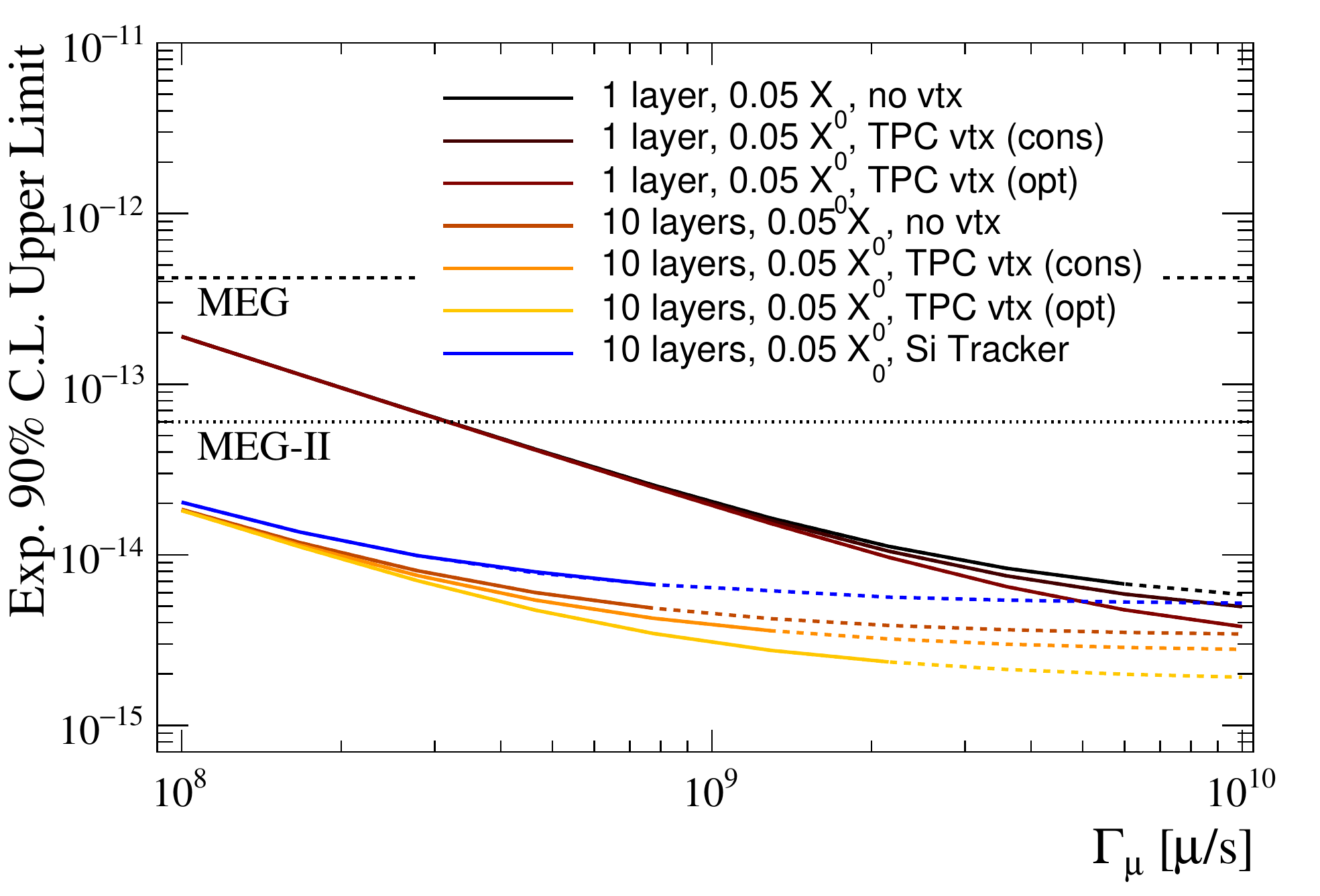}
\includegraphics[scale=.4]{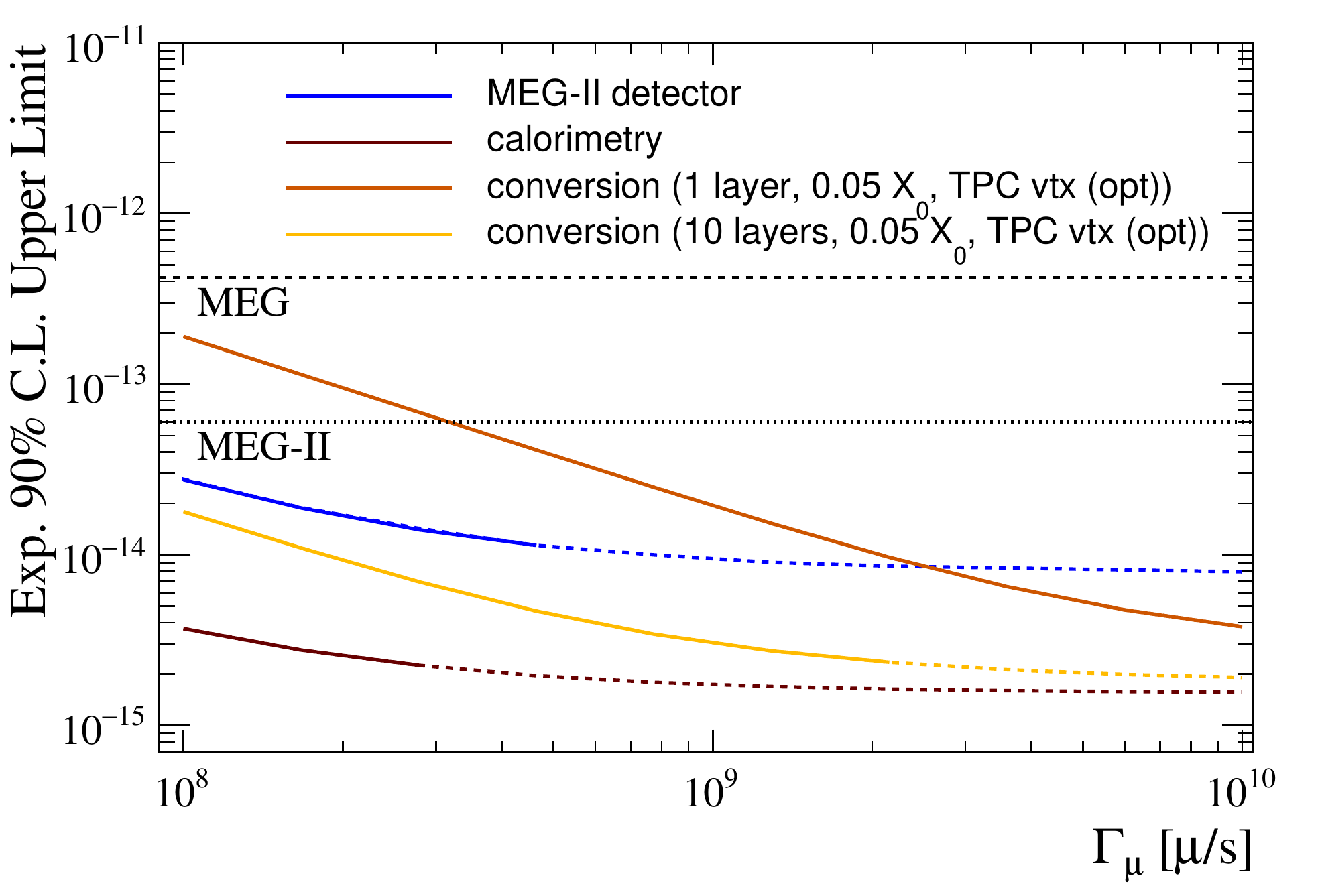}
\caption{\label{fig:next_meg}Projected sensitivity of a conceptual experiment searching for \meg~in a three year run, as a function of the muon beam intensity 
in different scenarios~\cite{meg-future}. Left: experiments with a photon conversion technique (one or more conversion layers), with and without a 
gaseous (Time Projection Chamber, TPC) or solid state (Silicon, Si) vertex detector. Right: comparison of sensitivities for experiments based on the photon 
conversion or calorimetric technique. The dashed line indicates scenarios where the number of accidental events over a three year run exceeds 10. The MEG 
and the expected MEG-II limits are also shown (dashed and dotted black line, respectively).}
\end{figure}

Studies have been performed to determine the possible sensitivity of future experiments searching for $\mu \to e \gamma$, in the view of the 
proposals to develop new high intensity muon beams, delivering a muon rate one or even two orders of magnitude larger than what is currently available.
At PSI the HiMB project~\cite{HiMB-1,HiMB-2} wants to exploit an optimized muon production target, a new system of solenoids to increase by a factor $> 4$
the efficiency of capturing the muon toward the beam line and a new design of the beam line optics to increase the transport efficiency by a factor $\sim 6$. The
goal is to reach a muon beam intensity from a few $10^{9}$ up to $10^{10}~\mu$/s. The main limitation to the achievable rate per incoming proton at PSI 
is the use of a thin production target to preserve the proton beam for a downstream neutron spallation source. The MuSIC project~\cite{MuSIC} at 
RCNP (Osaka, JP) wants to exploit a different technique, with a smaller proton current but a much thicker target, and a specifically designed capture solenoid.
The expected muon yield per unit of beam power is three orders of magnitude larger than what is obtained at PSI. Although the present expectations do not exceed
significantly the rates already available at PSI, it is an example of a new strategy for the production of intense muon beams. Finally, studies to produce continuous 
muon beams with rates exceeding $10^{9}~\mu$/s are also on going in the framework of the PIP-II project at Fermilab~\cite{PIP-II}. In~\cite{meg-future} we investigate
the experimental factors that will ultimately limit the sensitivity of a \meg~experiment with such large beam rates. While a magnetic spectrometer with gaseous
detectors still looks the best choice for the positron reconstruction, the higher rate tolerance of solid state detectors could make them preferable at the highest
intensities, although with a significative loss of momentum resolution. Anyway, the performances will be ultimately limited by the interaction of the positrons
in the detector material, and namely in the stopping target. It seems difficult, for instance, to reach angular resolutions below 3~mrad. On the photon side, the 
development of very luminous and fast scintillating crystals like LaBr$_{3}$(Ce) could be a good alternative to LXe. The cost of these materials, anyway, is still
very large and could limit the acceptance that can be reasonably reached. The photon conversion technique could be again a good alternative, in particular
if very high beam rates are available and a good timing can be obtained with active converters~\cite{TT-PET}. Taking into account all these limiting factors,
a few experimental scenarios have been considered, with calorimetry, photon conversion (one or more layers) and possibly the inclusion of a gaseous 
or solid state vertex detector to improve the vertex and angle resolutions. As shown in Figure~\ref{fig:next_meg}, an UL around a few $10^{-15}$ could be 
obtained with standard techniques but a non trivial R\&D effort, while breaking the $10^{-15}$ level would require a more radical rethinking of the 
experimental concept.
 
\subsection{\meee}
\label{sec:meee}

The \meee decay is at first the natural extension of the \meg~decay, when the photon is virtual and converts internally into $e^+e^-$. Anyway,
NP could also include an interaction between the four fermions without an intermediate photon. In the language of EFT, the \meg~decay at tree
level is produced by a dipole-like interaction,
\begin{equation}
\label{eq:dipole}
\mathcal{L}_\mathrm{dipole} = \frac{1}{\Lambda^2}C_L^D q_e m_\mu \, \left[\, \overline e \, \sigma^{\mu\nu}\frac{1-\gamma^5}{2} \mu \, \right] \, F_{\mu\nu} \; ,
\end{equation}
but four-fermion operators could also be present in the lagrangian, for instance:
\begin{equation}
\mathcal{L}_\mathrm{V,LL,e e} = \frac{1}{\Lambda^2}C^{V\,LL}_{ee} \left[\, \overline e \, \gamma^\mu\frac{1-\gamma^5}{2} \mu \, \right] \left[\, \overline e \, \gamma^\mu\frac{1-\gamma^5}{2} e \, \right]  \; .
\end{equation}
Anyway, the naive conclusion that the \meg~channel is only sensitive to dipole interactions while the \meee~channel also receives contributions from four-fermion
operators has to be disregarded. It has been shown~\cite{crivellin} that if all operators are consistently taken into account, the mixing introduced at loop level
by the renormalization-group evolution gives contributions to \meg~from four-fermion operators, and there are scenarios where the searches for \meg~are even more
sensitive than the searches for \meee~to constrain some of them. There is hence a strong complementarity between searches for cLFV in different channels.

The main experimental difference in \meee~with respect to \meg~comes from the presence of three charged particles whose direction can be precisely
measured in a tracking detector, and one can require that they come from the same point within the experimental resolutions. It drastically reduces
the likelihood of an accidental coincidence. The main background is thus a radiative muon decay where the photon is virtual and converts internally into
$e^+e^-$, i.e. $\mu^+ \to e^+ e^+ e^- \overline \nu_\mu \nu_e$, if the two neutrinos carry a very low fraction of the available energy. On the
other hand, the three-body decay kinematics imposes a detector design with a much larger momentum acceptance (compared for instance to MEG, where
only tracks with a momentum larger than 40~MeV could be detected), and the only kinematical discriminating variables are the vector sum of the track momenta and
the sum of the track energies, which have to be consistent with zero and with the muon mass, respectively.

The \meee decay was searched in the years 1980s by the Crystal Box experiment, which set the limit $BR(\meee) < 3.5 \times 10^{-11}$ at 90\% CL~\cite{CrystalBox}.
The best limit comes anyway from the SINDRUM experiment~\cite{sindrum}, operated at PSI in the same years. The experiment was a magnetic spectrometer 
instrumented with MWPCs and plastic scintillators for timing and trigger, surrounding a muon stopping target shaped as a double hollow cone. With such a 
design of the target, the beam is spread over a large surface, further reducing the probability of an accidental coincidence. It gives the best compromise between accidental 
background rejection, stopping efficiency and minimization of the material along the trajectory of the decay products. The momentum resolution of the 
spectrometer was 1.4 and 0.7~MeV/c in the component parallel and perpendicular to the decay plane, respectively. The experiment did
not observe any significative signal, and set the UL $BR(\meee) < 1.0 \times 10^{-12}$ at 90\% CL.

\begin{figure}[!t]
\centering
\vspace{-7cm}
\includegraphics[scale=.7]{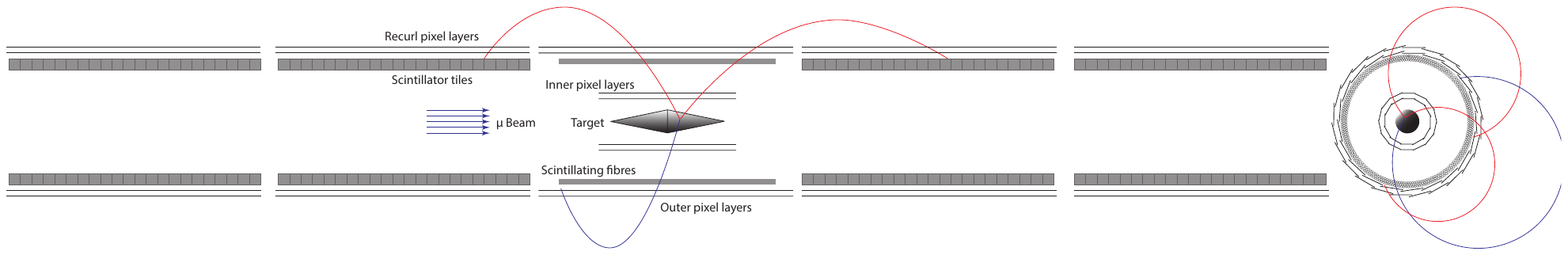}
\vspace{-9cm}
\caption{\label{fig:mu3e}A sketch of the Mu3e experiment with typical positron (red) and electron (blue) trajectories. Only the inner and outer 
pixel layers will be installed in Phase IA. In Phase IB the first two recurl stations and the scintillators (tiles and fibers) will be added. Two additional 
recurl stations will be installed in Phase II.}
\end{figure}

A new experiment to search for \meee~is under construction at PSI~\cite{mu3e-proposal}. The Mu3e detector, sketched in Figure~\ref{fig:mu3e}, 
will exploit the most recent developments in the construction of thin silicon detectors, adopting the HVMAPS technology~\cite{mu3e-hvmaps}. 
These devices, composed of a $50~\mu$m silicon pixel sensor on top of a light $100~\mu$m kapton structure including the power and signal routing, 
for a total of less than $10^{-3}$ radiation lengths, allow to keep a very low material budget, although the detector resolutions will be still limited by the 
multiple Coulomb scattering. The experiment will be developed in three phases. The Phase IA detector will be installed on the same beam line used for the 
MEG experiment, and will be composed of two double layers of silicon detectors, in a 1~T solenoidal magnetic field. A 100 day data taking with 
$2 \times 10^7~\mu$/s is expected to start around 2021. In Phase IB two stations of double silicon layers will be added (\emph{recurl stations}), 
providing an average resolution of 0.65~MeV on the sum of the track energies, and scintillators (fibers and tiles) will be included for a precise timing, 
so to exploit the maximum beam rate presently available, $\sim 10^8~\mu$/s, without being overcome by the accidental background. In Phase II the 
detector should be moved to the new high-intensity beam line discussed in the Sec.~\ref{sec:meg}, with additional tracking and timing stations and an 
average energy sum resolution of 0.52~MeV. The goal is to improve by a factor 20 the SINDRUM limit already in Phase IA, reach a sensitivity of $O(10^{-15})$ 
after Phase IB and push it down to $10^{-16}$ with $\sim 300$ days of run in Phase II at $2 \times 10^9~\mu$/s.

\subsection{$\mu \to e$ conversion in the Coulomb field of a nucleus}
\label{sec:mu2e}

A negative muon stopped on a target is captured by the Coulomb field of a nucleus and forms a muonic atom. 
In $\sim 10^{-16}$~s the muon cascades down to the $1s$ atomic level, where it can undergo a decay in orbit (DIO) or be 
captured by the nucleus through the process $\mu^- p \to \nu_\mu \, n$. The first case has to be treated
as the decay of a bound state, $(\mu^-,A(Z,N)) \to A(Z,N) \, e^- \nu_\mu \overline \nu_e$. Hence, with respect to the decay of a free muon, the
kinematical end point of the electron energy spectrum is not $m_\mu/2$ but, when the two neutrinos carry zero energy:
\begin{equation}
E_{0\nu} = \frac{M^2 - M^2_A}{2M} \; ,
\end{equation}
where $M = m_\mu + M_A - E_B$, being $M_A$ the mass of the nucleus and $E_B$ the binding energy of the system. For the
muonic Aluminum, $E_B = 0.48$~MeV and hence $E_\mathrm{max} = 104.96$~MeV.

If LFV is allowed, the muon can interact for instance with virtual photons from the Coulomb field of the nucleus through a dipole interaction like Eq.~(\ref{eq:dipole}), 
and decay into an electron of energy $E_{0\nu}$, without the emission of neutrinos, $(\mu^-,A(Z,N)) \to A(Z,N) \, e^-$. The process, 
usually referred to as $\mu \to e$ conversion in nuclei, can occur also through a four-fermion interaction and the same theoretical considerations 
of Sec.~\ref{sec:meee} apply.

From the experimental point of view, the search for $\mu \to e$ conversion consists in looking for electrons emitted with energy $E_{0\nu}$ from the 
muon stopping target, and it has to face several backgrounds. As already mentioned, the spectrum of electrons from the DIO extends up to $E_{0\nu}$,
although it quickly decreases near the end point as $(E - E_{0\nu})^5$. If the muon is captured by the nucleus, a high energy photon can be emitted and
convert into $e^+e^-$, where $e^-$ can have an energy near the end point. The same happens if pions contaminating the beam are captured in the target,
and, in general, particles in the beam other than muons can produce a background, as well as cosmic rays going through the detector (indeed, one
of the main sources of background in the next generation of experiments). One can consider prompt beam-related, delayed beam-related, physics
and cosmic backgrounds, according to the classification in Table~\ref{tab:mu2e-bkg}.

\begin{table*}[!h]
\caption{\label{tab:mu2e-bkg}Classification of backgrounds in $\mu \to e$ conversion experiments (table from~\cite{comet-cdr}).}
\centering
\begin{tabular}{|l|l|}
\hline
\multicolumn{2}{|c|}{Prompt beam-related backgrounds}\\
\hline
Radiative pion capture & $\pi^- + A \to \gamma + A^\prime$ and $\gamma \to e^+e^-$\\
Beam electrons & $e^-$ scattering of a muon stopping target\\
Muon decay in flight & $\mu^-$ decays in flight to produce $e^-$\\
Pion decay in flight & $\pi^-$ decays in flight to produce $e^-$\\
Neutron induced backgrounds & neutrons hit material to produce $e^-$\\
\hline	
\multicolumn{2}{|c|}{Delayed beam-related backgrounds} \\
\hline
Delayed-pion radiative capture & $\pi^- + A \to \gamma + A^\prime$ and $\gamma \to e^+e^-$\\
Anti-proton induced backgrounds & $\overline p$ hits material to produce $e^-$\\
\hline
\multicolumn{2}{|c|}{Physics backgrounds} \\
\hline
Muon decay in orbit & Decay of muons bound in a muonic atom \\
Radiative muon capture &  $\mu^- + A \to \nu_\mu + A^\prime$, $A^\prime \to \gamma + A$ and $\gamma \to e^+e^-$\\
$\mu^-$ capture with neutron emission & $\mu^- + A \to \nu_\mu + A^\prime$, $A^\prime \to n + A$ and neutrons produce $e^-$\\
$\mu^-$ capture with charged particle emission & $\mu^- + A \to \nu_\mu + A^\prime$, $A^\prime \to p(d,\alpha) + A$ and $p(d,\alpha)$ produce $e^-$\\
\hline
\multicolumn{2}{|c|}{Cosmic backgrounds} \\
\hline
Cosmic-ray induced backgrounds & \\
\hline
\end{tabular}
\end{table*}

\begin{figure}[!t]
\centering
\includegraphics[scale=0.7]{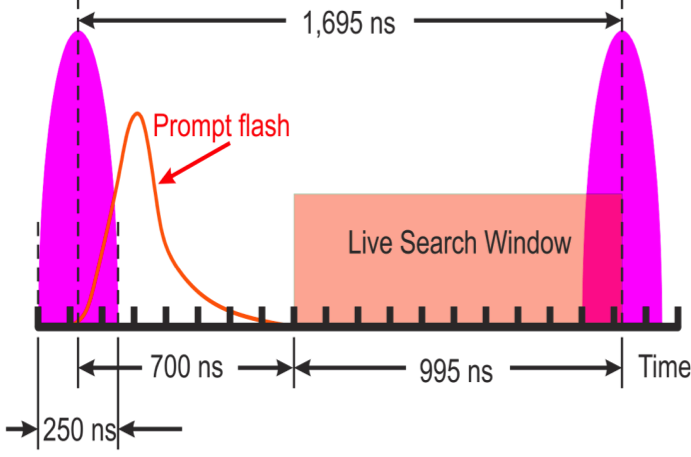}
\caption{\label{fig:mu2e-timing}Time structure of the beam and data acquisition timing scheme for the Mu2e experiment.}
\end{figure}

\begin{figure}[!t]
\centering
\vspace{-1cm}
\includegraphics[scale=0.4]{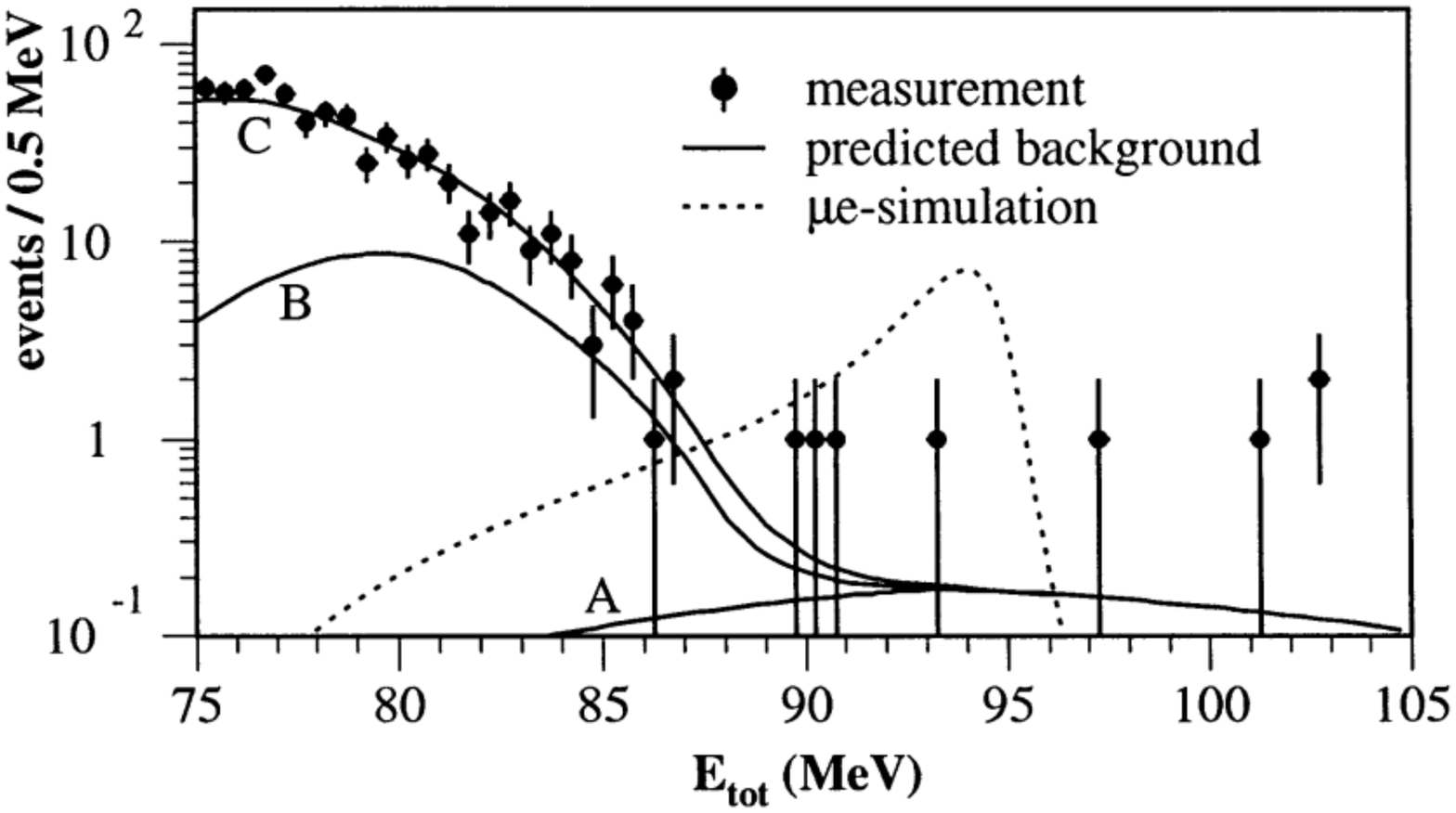}
\vspace{-1cm}
\caption{\label{fig:sindrum-lead} Electron momentum spectrum measured at SINDRUM-II with a Lead target~\cite{sindrum-Pb}.}
\end{figure}

The limits for the $\mu \to e$ conversion in nuclei are expressed in terms of the ratio between the conversion and nuclear capture rates:
\begin{equation} 
R_{e\mu} = \frac{\Gamma(\mu^- A(Z,N) \to e^- A(Z,N))}{\Gamma(\mu^- A(Z,N) \to \nu_\mu \, A(Z-1,N+1))} \, .
\end{equation}
The best limits currently available come from the SINDRUM-II experiment~\cite{sindrum-Ti} at PSI. The detector was a magnetic spectrometer
instrumented with two cylindrical drift chambers, with a momentum resolution of 2.3~MeV/c (FWHM). The experiment also included
a Cherenkov detector to identify the electrons and scintillators for trigger purposes. A continuous muon beam was used, delivering
about $\sim 10^7~\mu$/s. Data were taken with Lead, Gold and Titanium targets, and no excess of events was observed in the signal region, 
leading to the limits quoted in Tab.~\ref{tab:sindrum}. SINDRUM-II also searched for the process $\mu^- \mathrm{Ti} \to e^+ \mathrm{Ca}$, 
setting the limit $R_{e\mu} < 1.7 \times 10^{-12}$. As an example, figure~\ref{fig:sindrum-lead} shows the electron momentum 
spectrum measured at SINDRUM-II with the Lead target.

\begin{table*}[!h]
\caption{\label{tab:sindrum}90\% CL upper limits on $R_{e\mu}$ obtained by the Sindrum-II experiment for different targets.}
\centering
\begin{tabular}{|l|c|c|}
\hline
Target & $R_{e\mu}$ UL (90\% CL) & Reference \\
\hline
Titanium & $4.3 \times 10^{-12}$ & \cite{sindrum-Ti} \\
\hline
Gold & $7 \times 10^{-13}$ & \cite{sindrum-Au}\\
\hline
Lead & $4.6 \times 10^{-11}$ & \cite{sindrum-Pb}\\
\hline
\end{tabular}
\end{table*}

Three experiments currently aim to get better limits: Mu2e, COMET and DeeMe. In order to reduce the backgrounds deriving from particles other 
than muons, pulsed beams will be used in these experiments. Indeed, if the data acquisition time window is opened only after a few hundred ns from 
the arrival of the proton bunch, both the prompt and delayed beam-related backgrounds are strongly suppressed, and one expects no particle to 
reach the detector, apart from the decay products of the muonic atom. A schematic of the beam structure and acquisition timing is shown in 
Figure~\ref{fig:mu2e-timing} for the Mu2e experiment. Anyway, one of the main challenges of this generation of experiments is
the requirement of keeping below $10^{-10}$ the fraction of protons arriving to the muon production target in between two bunches 
(the \emph{extinction} of the beam), in order to make the beam-related backgrounds negligible.

\begin{figure}[!t]
\centering
\includegraphics[scale=.3]{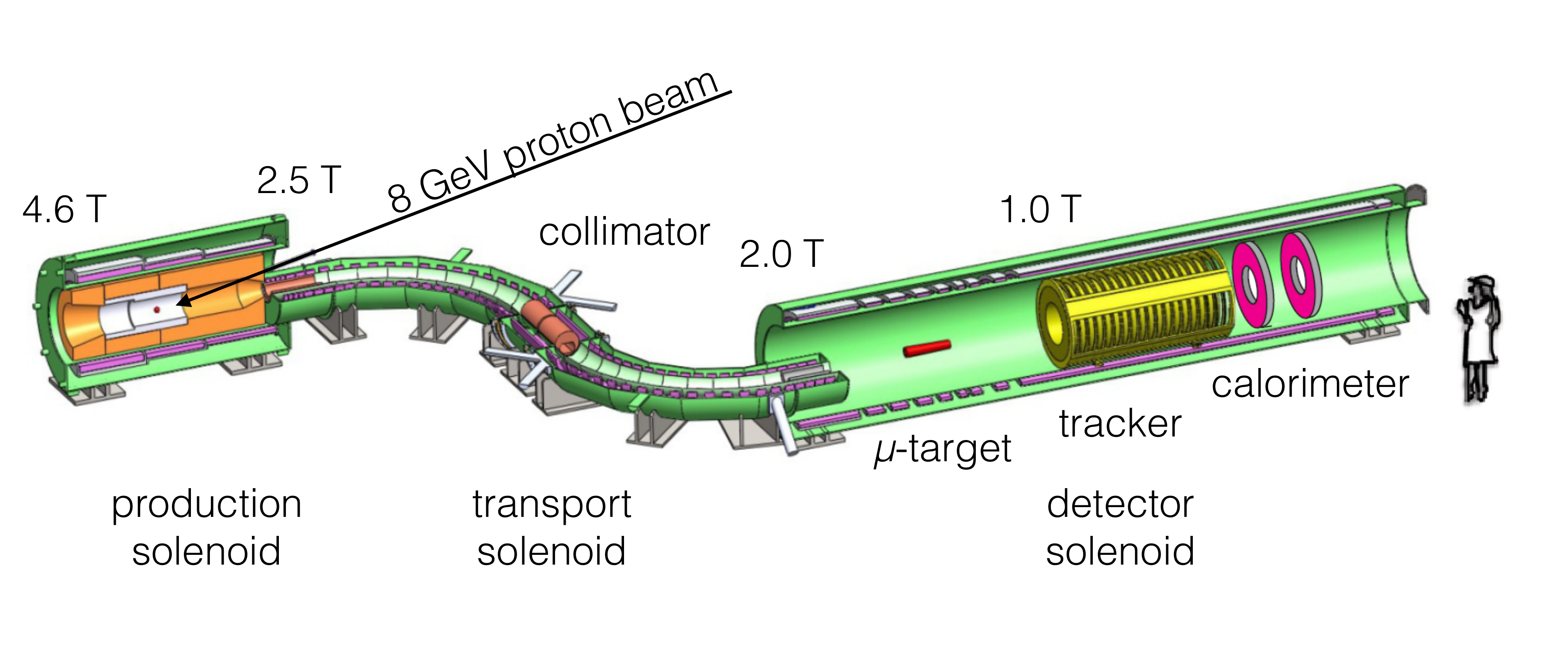}
\caption{\label{fig:mu2e}A sketch of the Mu2e experiment.}
\end{figure}

The Mu2e experiment~\cite{mu2e} is under construction at Fermilab, and is sketched in Figure~\ref{fig:mu2e}. Protons of 8~GeV, in bunches of 200~ns separated
by 1695~ns, hit a target immersed in a graded magnetic field going from 2.5 to 4.6~T along the incoming beam direction. Slow pions produced in the 
target tend to be reflected by the magnetic field in the opposite direction with respect to the incoming protons, and the low momentum muons 
produced in their decay enter a curved solenoid magnet. In this magnet the muons spiralize around the field direction and are transported 
toward the muon stopping target. The curved solenoid prevents the passage of neutral particles and, thanks to the field gradients which are necessarily 
present in the curved sectors, allows to separate positive and negative particles, so that the formers can be removed using collimators. The muons finally reach the 
stopping target, a set of seventeen $0.02~\mu$m Aluminum foils (using multiple thin foils allow to reduce the energy loss of the electrons exiting the target while 
keeping a large stopping power). Aluminum have been chosen as the best compromise between stopping efficiency, material budget seen by the electrons and
muonic atom lifetime (880~ns, long enough to wait for prompt beam backgrounds to be strongly suppressed before opening the acquisition window).
The magnetic field in the target region also has a gradient from 2 to 1~T, to maximize the detector acceptance by reflecting a fraction of the back-emitted electrons.
The momentum of the electrons produced in the target is measured by a system of 20,000 drift tubes in a 1~T solenoidal magnetic field downstream of the stopping
target, with a resolution of 300~keV/c. Then the electrons reach a calorimeter made of CsI crystals read out by silicon photon detectors (SiPM). This detector 
allows to distinguish electrons from muons and helps in the track reconstruction. A large set of scintillators to veto cosmic ray events will cover the whole detector. 
The experiment is expected to be built by 2021 and take data for three years (2022-2025) with $3.6 \times 10^{20}$ protons on target, $6 \times 10^{17}$ stopped 
muons and 0.46 expected background events. The expected limit is $R_{e\mu} < 8 \times 10^{-17}$~\cite{mu2e-2017}.

\begin{figure}[!t]
\centering
\includegraphics[scale=.3]{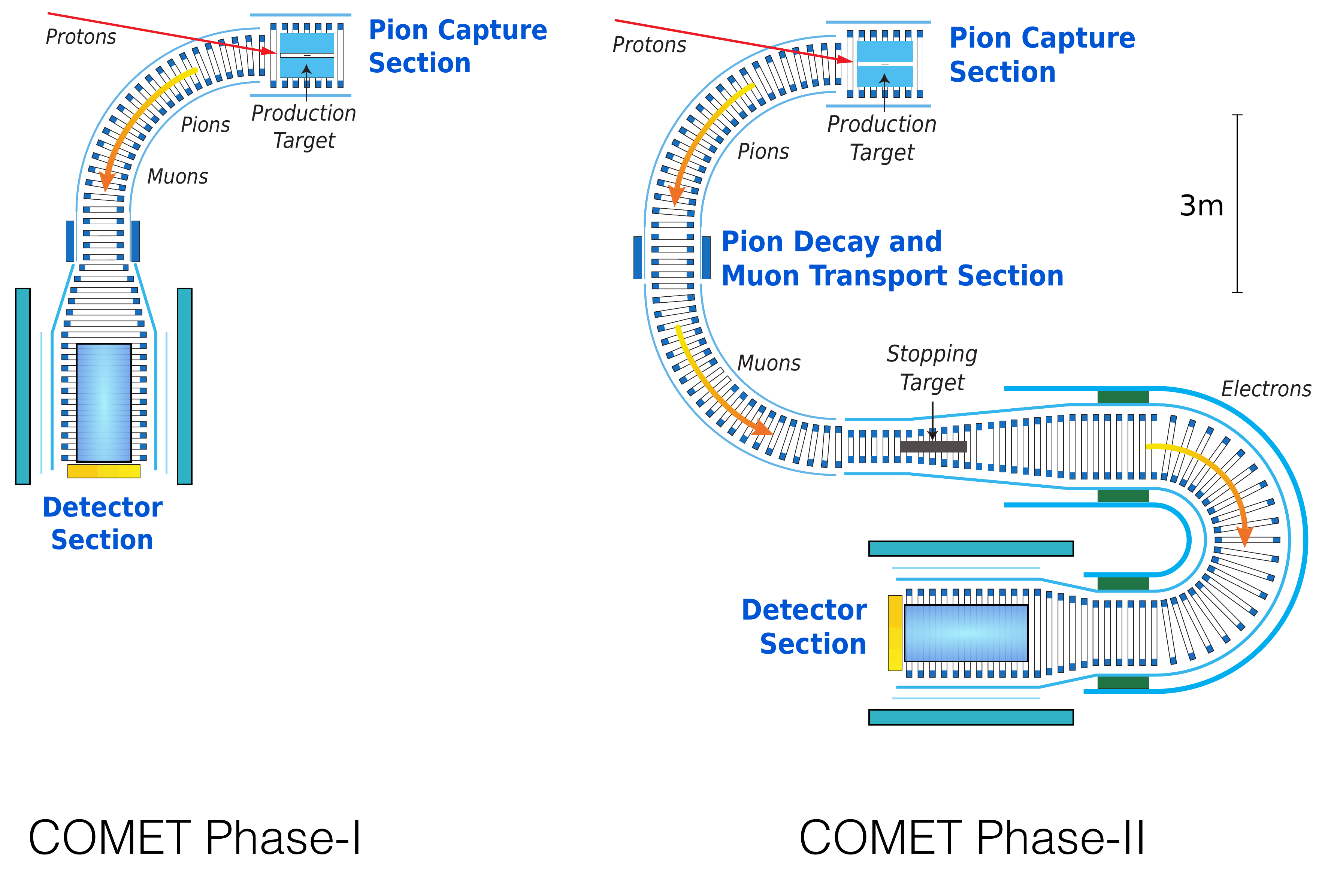}
\caption{\label{fig:comet}A sketch of the COMET experiment.}
\end{figure}

The concept of the COMET experiment~\cite{comet} at J-PARC (Tokai, JP) will be very similar in its first phase. A pion capture solenoid with a graded field from 5 to 2~T
will be followed by a 2~T transport solenoid, curved by $90^\circ$, a target stopping immersed in a 3 to 1~T graded field and a 
detector section with a 1~T field. The detector will be composed of a set of 5 planes of drift tubes, for a total of 4160 detectors, followed by a Gd$_2$SiO$_5$ 
calorimeter. A momentum resolution of 200~keV/c is expected. In the second phase, the setup will include a second section of the transport solenoid, 
thus making a C shape, and a second C-shaped solenoid between the stopping target and the detector. The former will improve the particle separation in the muon
beam, the latter will reduce the event rate in the tracker by preventing low momentum electrons to reach the detector, and will efficiently remove the 
proton background. The setup for the two phases is shown in Figure~\ref{fig:comet}. The experiment is expected to start its first phase in 2019, collecting up to 
$\sim 10^{16}$ stopped muons and reaching an UL of $\sim 7 \times 10^{-15}$ on $R_{e\mu}$ after two years. The second phase will push the limit down to 
$< 10^{-16}$ with a two-year data taking ($\sim 2 \times 10^{18}$ stopped muons) starting in 2022.

\begin{figure}[!t]
\centering
\includegraphics[scale=.25]{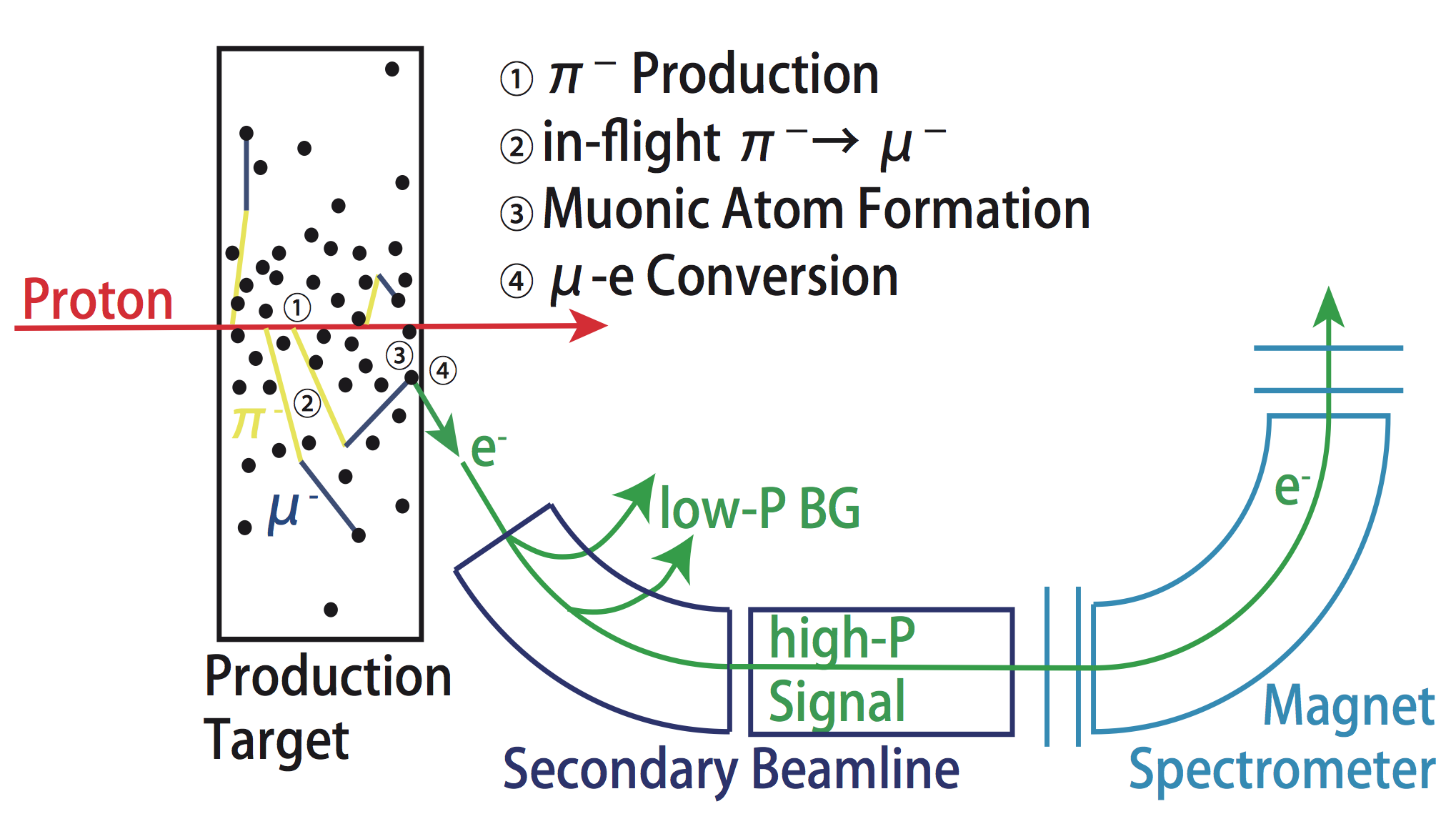}
\caption{\label{fig:deeme}A sketch of the DeeMe experiment.}
\end{figure}

The DeeMe experiment~\cite{DeeMe}, also operated at J-PARC, aims to reach a less ambitious limit, but in a shorter time scale. In this experiment, sketched in Figure~\ref{fig:deeme}, muons
are directly captured in a silicon-carbide proton target and the emerging electrons are transported by a beam line toward a dipole magnet preceded an followed by pairs of
MWPCs, so that their momentum can be measured. The beam line allows to reject low-energy and positive-charged particles. 
The experiment is expected to start taking data soon and reach a limit $O(10^{-14})$ in a four-year run.

\section{Muon decays to exotic particles}
\label{sec:exotics}

In this section we consider the production of new particles in muon decays. We consider in particular scenarios where a boson $X$ of mass $m_X < m_\mu - m_e$ is
emitted in $\mu^+$ decays in association with a positron and, possibly, an irradiated photon. Such bosons typically emerge as goldstone or pseudo-goldstone bosons 
in models with respectively exact or approximate symmetries that are spontaneously broken. In the first case the particle is massless, in the second case it
is massive. The list of such hypothetical particles includes axions and axion-like particles, majorons, familons, light Z' bosons, dark photons etc. 
A recent and rich bibliography on this subject can be found in~\cite{heeck}. It is also worth mentioning that the interest on light bosons in the mass region 
that can be explored with muon decays has been also recently boosted by an anomaly emerged in the angular spectrum of $e^+e^-$ pairs emitted in the 
de-excitation of a $^8\mathrm{Be}^*$ state, which could be interpreted as the production of a new boson of mass $\sim 16.7$~MeV~\cite{atomki}, possibly 
the carrier of a new fundamental force~\cite{fifth}.

From the experimental point of view, inclusive searches have been mostly performed. In the two body decay $\mu^+ \to e^+ X$, monoenergetic positrons would
be emitted, with $E_e = m_\mu/2$ in the massless case, and experiments look for a peak in the momentum spectrum of the positrons emerging from 
a muon stopping target. If the $\mu^+ \to e^+ X \gamma$ process is considered, events with a definite missing invariant mass 
$m_X^2 = (p_\mu - p_e - p_\gamma)^2$ are searched for. These searches allow to look for both stable or unstable particles. 
Exclusive searches are also possible if the boson decays into electrons and photons, by fully reconstructing its decay products. 
If for instance the muon decays in the channel $\mu^+ \to e^+ X$ with $X \to e^+ e^-$ or $X \to \gamma \gamma$, one can look for a peak 
in the distribution of the $e^+e^-$ or $\gamma\gamma$ invariant mass. If there is no requirement on the positron kinematics, 
these searches are also sensitive to the lepton-flavor-conserving channel $\mu^+ \to e^+ \overline \nu_\mu \nu_e X$.
This search is of particular interest in the framework of models with dark photons~\cite{darkphoton}, in which case this is a RMD where a dark photon is irradiated
in place of a normal photon.

\subsection{Inclusive searches}

Inclusive searches for a new boson in the $\mu^+ \to e^+ X$ decay are most profitably performed by experiments studying the positron momentum spectrum
of Michel decays, because these experiments are typically designed to have a large momentum acceptance and provide a very precise and accurate determination
of the positron momentum. The first feature allows to look at a wide range of $X$ masses, the second one provides optimal conditions to look for
monoenergetic peaks. The momentum accuracy is of paramount importance in the case of a massless boson, when positrons with energies at the 
kinematical end point $E_e = m_\mu/2$ are expected. The Michel spectrum has a steep edge in this region and, as a result, an unrecognized bias in the measurement 
of the momentum, i.e. a mismatch between the expected and measured position of the edge, can easily mimic the presence of a peak at the end point superimposed to
the normal spectrum.

The most recent limits for the search of massless and massive particles in $\mu^+ \to e^+ X$ come from the TWIST experiment~\cite{twist-x}. 
A significative effort was made to reduce the systematic uncertainties related to the calibration of the momentum reconstruction. 
A resolution of (58~keV/c)/$\sin\theta_e$ with an accuracy $\sim 6$~keV/c could be finally reached. 
The BR of the decay $\mu^+ \to e^+ X$ with massive or massless $X$ is constrained, assuming either an isotropic $e^+$ angular distribution or 
anisotropies induced by the muon polarization. An average UL of $8.1 \times 10^{-6}$ at 90\% CL for masses from 10 to more than 80~MeV/c$^2$ is obtained, 
while the limit $\mathrm{BR}(\mu^+ \to e^+ X) < 3.3 \times 10^{-5}$ is extracted for a 
massless $X$. A better limit was obtained in the past by Jodidio \emph{et al.} for a massless $X$, 
$\mathrm{BR}(\mu^+ \to e^+ X) < 2.5 \times 10^{-6}$~\cite{jodidio}. The angular acceptance of this
experiment was limited anyway to the forward region with respect to the muon polarization, and hence the limit is not valid if the
positrons are emitted prevalently in the opposite direction with respect to the muon spin.

It should be noticed that experiments searching for \meg~can also be sensitive to the $\mu^+ \to e^+ X$ channel, even if their trigger is optimized
to search for correlated pairs of a positron and a photon. Indeed, as mentioned in Sec.~\ref{sec:meg}, most of the events collected by these
experiments when running at high beam rates are accidental coincidences of uncorrelated positrons and photons, and the observed positron
momentum spectrum is in practice the Michel spectrum, on top of which one can look for a monoenergetic peak. Anyway, these experiments
are typically designed to reconstruct only the high-energy positrons (which limits the mass range that can be investigated) and the control
of biases in the reconstructed momentum, though good enough for the search of $\mu^+ \to e^+ \gamma$, is poorer with respect to
experiments, like TWIST, specifically designed to measure the Michel spectrum, giving non-competitive systematic uncertainties in the search 
for a massless boson, unless a specific effort is performed in this direction.

Conversely, \meg~experiments are the best place to look for $\mu^+ \to e^+ X \gamma$, thanks to their very good photon reconstruction,
although the trigger, which selects preferably photons and positrons emitted in opposite directions, limits the available phase space. The 
backgrounds for this search are RMDs and accidental coincidences, and the signal is looked for in the distribution of the missing invariant mass. 
The best limit currently available, $\mathrm{BR}(\mu^+ \to e^+ X \gamma) < 1.3 \times 10^{-9}$ at 90\% CL, has been produced by the 
Crystal Box collaboration~\cite{crystalbox-x}, using the same data set of their search for \meg. Although a dedicated study have not been performed yet,
the limit could be presumably improved by MEG-II.

The decay $\mu^+ \to e^+ X$ has been also searched for with a HPGe detector at PSI~\cite{HPGe}, for masses in the range $103 < m_X < 150$~MeV/c$^2$,
setting an upper limit $BR(\mu^e \to e^+ X) < 5.7 \times 10^{-4}$ at 90\% CL. In this experiment, the HPGe is used both as
a muon stopping target and as a detector, where the positron energy is deposited and measured.

\subsection{Exclusive searches for muon decays to unstable exotic particles}

If the new boson to be searched for is unstable and decays inside the detector into detectable standard particles (electrons and photons), 
an exclusive search can be performed, provided that its lifetime is short enough to have a significative fraction of events where
the particle decays inside the detector.

The $\mu^+ \to e^+ X$ decay with $X \to e^+ e^-$ can be investigated by experiments designed 
for \meee, looking for a peak in the distribution of the invariant mass of each $e^+e^-$ pair. Such a kind of analysis has been performed 
at SINDRUM, setting limits in a wide range of masses, including also the $\mu^+ \to e^+ X \nu$ decay mode~\cite{sindrum-x}. 
Limits down to $\mathrm{BR}(\mu^+ \to e^+ X) \times \mathrm{BR}(X \to e^+ e^-) < 2 \times 10^{-12}$ have been obtained for lifetimes below $10^{-10}$~s.

The sensitivity of the Mu3e experiment to light bosons in the channel $\mu^+ \to e^+ \overline \nu_\mu \nu_e X$ has been investigated~\cite{mu3e-dark}.
In Phase II, upper limits down to $O(10^{-11})$ on the BR of this process can be reached for a mass of $\sim 10$~MeV, for which the experiment 
would be maximally sensitive.

For what concerns $X \to \gamma \gamma$, the aforementioned search for $\mu \to e \gamma \gamma$ by Crystal Box~\cite{CrystalBox} cannot
be considered because it did not include any requirement on the $\gamma\gamma$ invariant mass. Preliminary studies have been performed at 
MEG~\cite{natori-tesi} and are planned at MEG-II, although this search will be mostly sensitive to small masses. Indeed, the angular acceptance of the 
LXe calorimeter of MEG and MEG-II is relatively small ($\Delta\phi_\gamma \sim 130^\circ$, $\Delta\theta_\gamma \sim 50^\circ$), and only photon 
pairs emitted with a small relative angle can be detected, corresponding to events where the decaying particle has a large Lorentz boost, which implies a 
low mass. The expected limits range from $O(10^{-11})$ for $M_X \sim 10$~MeV/c$^2$ to $O(10^{-7})$ at $M_X \sim 45$~MeV/c$^2$.

\section{Conclusions}

In this paper I reviewed the current status of the searches for physics beyond the SM in muon decays. Muon decays provide a unique environment 
to search for new physics, with an excellent sensitivity to a wide range of models and negligible theoretical uncertainties. Measurements of the differential
decay rates in $\mu^+ \to e^+ \overline \nu_\mu \nu_e$ and $\mu^+ \to e^+ \overline \nu_\mu \nu_e \gamma$ currently provide some
of the best tests of the V-A structure of weak interactions. A worldwide effort is devoted to the search of lepton-flavor-violating muon decays, with recent
results on \meg~and upcoming improvements by orders of magnitudes in the sensitivity to \meee, $\mu \to e$ conversion and \meg~itself. Experiments
designed to perform these searches are also sensitive to muon decays where new particles like Goldstone and pseudo-Goldstone bosons are produced.
The muon, the first particle to be studied with modern particle physics techniques in the pioneering work by Conversi, Pancini and Piccioni~\cite{CPP1},
still provides one of the best playgrounds for the development of fundamental physics.

\bibliographystyle{unsrt}
\bibliography{revip-muon}

\end{document}